\documentclass[fleqn,usenatbib]{mnras}

\usepackage{newtxtext,newtxmath}

\usepackage[T1]{fontenc}

\usepackage{multirow}
\usepackage{multicol}
\usepackage{makecell}
\usepackage{graphicx}
\usepackage[flushleft]{threeparttable}
\usepackage{booktabs}
\usepackage{comment}
\usepackage{hyperref}
\DeclareRobustCommand{\VAN}[3]{#2}
\let\VANthebibliography\thebibliography
\def\thebibliography{\DeclareRobustCommand{\VAN}[3]{##3}\VANthebibliography}

\usepackage{graphicx}	
\usepackage{amsmath}	
\usepackage{lscape} 

\title[Methyl cyanide gas-phase formation network] {Revised gas-phase formation network of methyl cyanide: the origin of methyl cyanide and methanol abundance correlation in hot corinos}

\author[Giani et al.]{
Lisa Giani,$^{1,2}$\thanks{E-mail:lisa.giani@univ-grenoble-alpes.fr}
Cecilia Ceccarelli$^{1}$,
Luca Mancini$^{2}$,
Eleonora Bianchi$^{3}$,
Fernando Pirani$^{2,4}$,
Marzio Rosi$^{4}$,
\newauthor Nadia Balucani$^{1,2}$
\\\\
$^{1}$Univ. Grenoble Alpes, CNRS, IPAG, 38000 Grenoble, France\\
$^{2}$Dipartimento di Chimica, Biologia e Biotecnologie, Universita degli Studi di Perugia, Perugia, 06123, Italy\\
$^{3}$ Excellence Cluster ORIGINS, Boltzmannstraße 2, 85748 Garching bei München, Germany\\
$^{4}$Dipartimento di Ingegneria Civile e Ambientale, Universita degli Studi di Perugia, Perugia, 06125, Italy\\
}

\date{Accepted XXX. Received YYY; in original form ZZZ}

\pubyear{2023}

\begin{document}
\label{firstpage}
\pagerange{\pageref{firstpage}--\pageref{lastpage}}
\maketitle

\begin{abstract}
Methyl cyanide (CH$_3$CN) is one of the most abundant and widely spread interstellar complex organic molecules (iCOMs).
Several studies found that, in hot corinos, methyl cyanide and methanol abundances are correlated suggesting a chemical link, often interpreted as a synthesis of them on the interstellar grain surfaces.
In this article, we present a revised network of the reactions forming methyl cyanide in the gas-phase.
We carried out an exhaustive review of the gas-phase CH$_3$CN formation routes, propose two new reactions and performed new quantum mechanics computations of several reactions.
We found that 13 of the 15 reactions reported in the databases KIDA and UDfA have incorrect products and/or rate constants.
The new corrected reaction network contains 10 reactions leading to methyl cyanide.
We tested the relative importance of those reactions in forming CH$_3$CN using our astrochemical model.
We confirm that the radiative association of CH${_3}{^+}$ and HCN, forming CH$_{3}$CNH$^{+}$, followed by the electron recombination of CH$_{3}$CNH$^{+}$, is the most important CH$_3$CN formation route in both cold and warm environments, notwithstanding that we significantly corrected the rate constants and products of both reactions.
The two newly proposed reactions play an important role in warm environments.
Finally, we found a very good agreement between the CH$_3$CN predicted abundances with those measured in cold ($\sim$10 K) and warm ($\sim$90 K) objects.
Unexpectedly, we also found a chemical link between methanol and methyl cyanide via the CH$_{3}^{+}$ ion, which can explain the observed correlation between the CH$_3$OH and CH$_3$CN abundances measured in hot corinos.  
\end{abstract}

\begin{keywords}
astrochemistry -- ISM: abundances -- ISM: molecules 
\end{keywords}



\section{Introduction} \label{sec:introduction}

Nitriles, that is, organic molecules with a -C$\equiv$N (cyano) functional group, comprise a considerable percentage ($\sim$15\%) of the almost 300 species detected in the interstellar medium (ISM) \citep[https://cdms.astro.uni-koeln.de/classic/molecules and][]{McGuire2022-ISMcensus}. 
They are key intermediates for the formation of biologically relevant molecules and, therefore, are believed to play a central role in prebiotic chemistry and the origin of life \citep[e.g.][]{Sanchez1966-HCNlife, das2019insights, meisner2019computational}. 
For instance, nitriles can hydrolyze and participate in multi-step synthesis of amino acids \citep{brack1998molecular} or RNA precursors \citep{powner2009synthesis,sutherland2017opinion}. 

Nitriles larger than HCN belong to the class of the so-called interstellar complex organic molecules (iCOMs), that is organic molecules containing at least 6 atoms \citep{herbst2009complex,ceccarelli2017seeds}. 
Methyl cyanide (CH$_{3}$CN, also known as acetonitrile) is the simplest organic nitrile and is one of the first detected \citep{Solomon1971-SgrB2-CH3CN} and most spread iCOMs. 
It has been detected in many galactic objects, remarkably in cold starless and prestellar cores \citep[e.g.][]{Minh1993-TMC1,Vastel2019, megias2023-PSCch3cn}, low-mass protostars and hot corinos \citep[e.g.][]{bottinelli2004near, bottinelli2004complex, taquet2015constraining, agundez2019sensitive, belloche2020, nazari2021complex, Nazari2022-NiCOMs, calcutt2018almanitriles, mercimek2022chemical, bianchi2022ch3cn, hsieh2023prodige} and protoplanetary disks \citep{oberg2015comet, ilee2021molecules}. 
Methyl cyanide has also been observed in comets \citep{ulich1974detection, rodgers2001organic, le2015inventory,altwegg2019cometary, biver2022observations}, and is one of the nitriles detected in Titan's atmosphere \citep[e.g.][]{Bezard1993-ch3cnTitan, marten2002new}. 
Therefore, CH$_{3}$CN is one of the few prebiotic species traced through all the formation phases of stars and planetary systems. 

As for other iCOMs, the formation route of methyl cyanide is debated \citep[e.g.][]{Ceccarelli2023-PP7, garrod2022formation, megias2023-PSCch3cn} and two major pathways have been evoked in the literature: either gas-phase or grain-surface reactions.
In the gas-phase, the sequence of the radiative association of CH${_3}{^+}$ and HCN leading to CH$_{3}$CNH$^{+}$ followed by the electron recombination of the protonated methyl cyanide would be the major pathway according to several astrochemical models \citep{Agundez2013-ChRv, garrod2022formation}. 
However, both steps have products and rate constants which have been discussed in the literature but with different conclusions \citep{mcewan1980low,anicich1993evaluated,bates1983theory,herbst1985update,defrees1985theoretical,klippenstein1996theory, anicich1995association, defrees1985theoretical} (see the detailed discussion in Sec. \ref{sec:overview-gas-routes}).

On the grain-surface, methyl cyanide would be formed either by the ice-assisted association of the CH$_3$ and CN radicals or by the hydrogenation of C$_{2}$N \citep{garrod2008complex, garrod2022formation}. 
Unfortunately, no experimental or computational data are available for these two processes. 
In the case of the CH$_3$ + CN recombination reaction, however, there might be energy barriers because of the interaction of the radicals with water-ice molecules, as observed in similar cases \citep{Enrique-Romero2019,Enrique-Romero2021,Enrique-Romero2022}. 
More important than that, it has been demonstrated that the CN radical readily reacts with the water molecules of the ice \citep{rimola2018can}, which makes recombination reactions involving CN radicals on ice highly unlikely.

Astronomical observations can be extremely useful to distinguish which of the two mechanisms prevails and in which environment.
For example, the observed correlation between the methyl formate and dimethyl ether abundances has inspired the work by \citet{balucani2015formation}, who proposed a chain of gas-phase reactions linking the two species.
Likewise, the observed correlation between ethanol, glycolaldehyde and acetaldehyde abundances has inspired the idea of a chain of gas-phase reactions starting from ethanol, formed on the grain-surfaces \citep[e.g.][]{Perrero2022-ethanol}, and producing other iCOMs \citep{skouteris2018genealogical}.
Although an observed correlation does not necessarily imply a direct chemical link \citep[e.g.][]{belloche2020}, it is worth to consider it.

In this context, a tight correlation between the CH$_{3}$OH and CH$_{3}$CN column density has recently been observed in a study of several low-mass protostars, the PEACHES survey \citep{yang2021perseus}.
The authors of this study privileged the interpretation that both CH$_{3}$OH and CH$_{3}$CN are formed on the grain surfaces, since this is the only known route of methanol formation.
However, no study has been explicitly carried out to verify whether gas-phase reactions may lead to the observed correlation, regardless whether due to a direct link between the two species.
Finally, the methyl cyanide deuteration ratio, [CH$_{2}$DCN]/[CH$_3$CN], measured towards the low-mass SVS13-A suggests that, in this source, methyl cyanide was synthesized in the gas-phase during the cold prestellar phase, condensed on the grain mantles and, eventually, injected into the gas-phase when the dust temperature reaches the mantles sublimation conditions, upon the heating from the central forming star \citep{bianchi2022ch3cn}. 


In this work, we focus on the gas-phase reactions leading to the formation of methyl cyanide.
To this end, we first carried out a critical search of the literature, starting from the list of reactions reported in the two astrochemical databases KIDA \citep[Kinetic Database for Astrochemistry:][]{wakelam2012kinetic} and UDfA \citep[UMIST Database for Astrochemistry:][]{mcelroy2013umist}.
Only a few of them have been investigated via laboratory experiments under the appropriate experimental conditions. 
In the cases in which the information was not complete, we performed dedicated quantum mechanics (QM) computations and estimated the relevant products and rate constants as a function of the temperature. 
In addition, we propose two new reactions which may contribute to the CH$_3$CN formation.
Also for these reactions, we carried out dedicated QM computations of the energetics and kinetics.
Finally, we obtained astrochemical model predictions using the new chemical network and compared them with the observations of methyl cyanide in cold and warm sources.




The article is organized as follows. 
Sec. \ref{sec:overview-gas-routes}, we present the critical review of the CH$_{3}$CN gas-phase formation reactions. 
The theoretical method for the QM computations is described in Sec. \ref{sec:methods}, while the results of the studied reactions, namely the potential energy surface (PES) and rate constants, are reported in Sec. \ref{sec:results}. 
The revised network is summarised in Sec. \ref{sec:revised-network} and the results obtained using the new network in the astrochemical model are shown in Sec. \ref{sec:astro-modeling}. 
We discuss all the results in Sec. \ref{sec:discussion}. 
Finally, the conclusions are given in Sec. \ref{sec:conclusions}.

\begin{figure}
	\includegraphics[width=8.5cm]{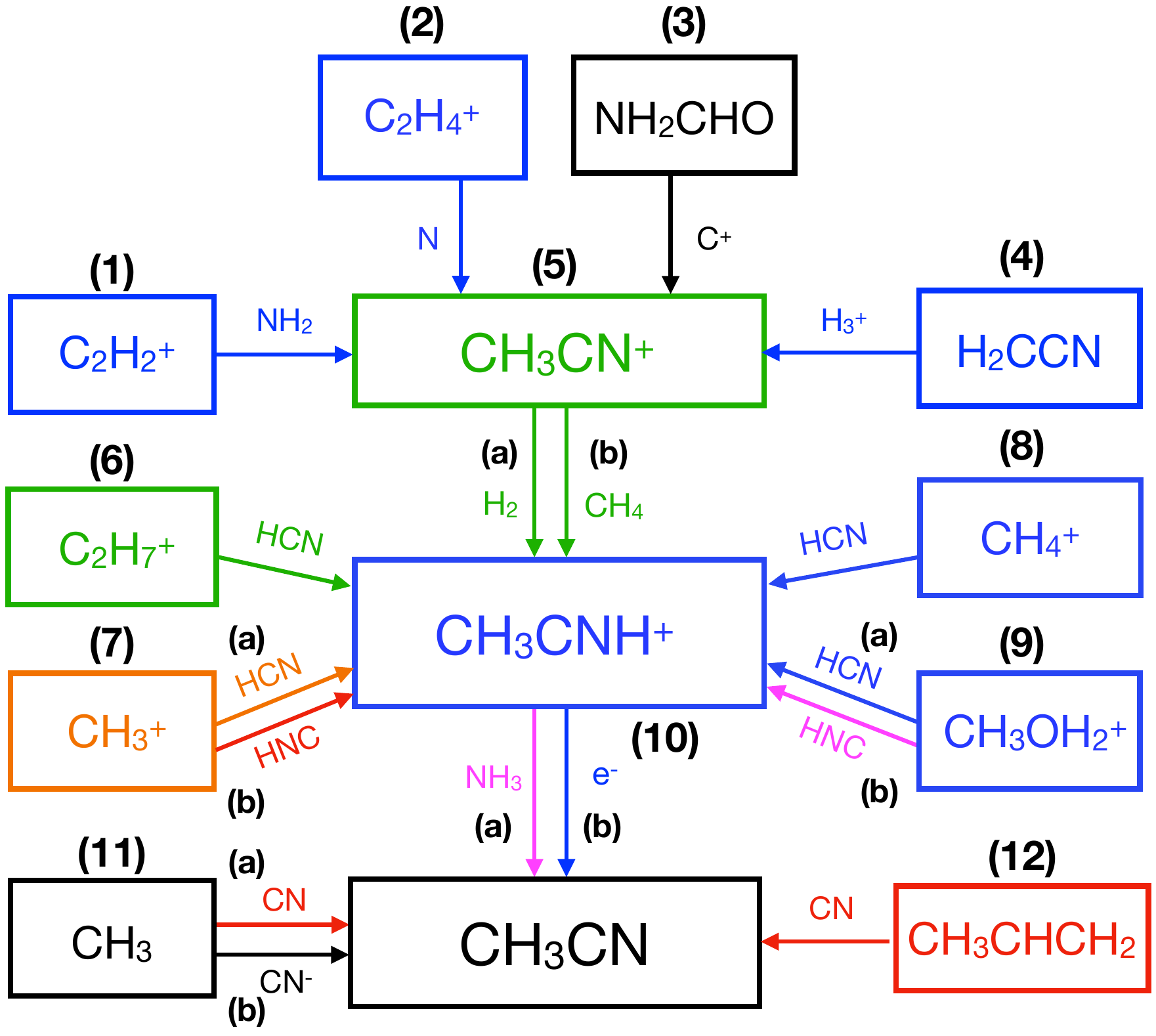}
    \caption{Scheme of the gas-phase reactions involved in the formation of methyl cyanide and listed in Table \ref{tab:overview-reactionlist}.
    The numbers on top of each box are those in Table \ref{tab:overview-reactionlist} and used in the subsections of Sec. \ref{sec:overview-gas-routes}.
    The color code of the boxes and arrows is the following:
    \textit{Green}, reactions correctly reported in the literature/databases;
    \textit{Blue}, reactions for which we have carried out new QM computations;
    \textit{Red}, reactions which we suggest removing from the network;
    \textit{Orange}, reactions which are updated with a different rate constant derived from the literature;
    \textit{Black}, reactions not revised in this work;
    \textit{Magenta}, newly proposed reactions. }
    \label{fig:overview-reactionscheme}
\end{figure}

\section{Critical overview of the gas-phase reactions forming methyl cyanide}\label{sec:overview-gas-routes}

Several gas-phase reactions have been proposed in the literature for the formation of methyl cyanide, most of which are included in the two commonly used astrochemical databases, KIDA and UDfA. 
However, often rate constants and product branching ratios (BR) have only been guessed on the basis of similar reactions rather than determined in experimental or theoretical studies.
Here, we provide a critical review of the reactions considered so far as contributing to the formation of methyl cyanide. 

\begin{table*}
    \centering
    \caption{List of the gas-phase reactions evoked in the literature, present in the KIDA or UDfA databases, and proposed in this work leading to the formation of methyl cyanide.
    The table is divided in three parts, according to which precursor they form: CH$_3$CN$^+$, CH$_3$CNH$^+$ or directly CH$_3$CN, respectively, following the scheme of Fig. \ref{fig:overview-reactionscheme}.
    The numbers in the \textit{first column} correspond to those also marked in Fig. \ref{fig:overview-reactionscheme} and used in the subsections of Sec. \ref{sec:overview-gas-routes}.
    \textit{Second and third columns} report the reactants and products of the reaction, respectively.
    The \textit{fourth column} reports the conclusions of our critical review of the literature for each reaction: "New QM computations" means that we carried out new computations; "Further investigation needed" means that the values (products and/or reaction constants) reported in the databases are not reliable but the reaction does not seem to be important for the CH$_3$CN formation (based on the astrochemical modeling of Sec. \ref{sec:astro-modeling}), so we did not carry out new computations; "Reliable" means that the values in the databases are correct; "Modified" means that the values have been updated based on the literature search; "Removed" means that the reaction has been removed because the CH$_3$CN is not a product of the reaction.
    The \textit{fifth column} reports the results of the new QM computations: "Removed" if the reaction has been removed because the CH$_3$CN is not a product of the reaction; "Modified" if the values of the reaction were modified; "Added" if the reaction has been added to the network.
    }
    \label{tab:overview-reactionlist}
    \begin{center}
    \begin{tabular}{llclll}
    \hline
    No. & Reactants & & Products & Review conclusions & QM computation results\\
    \hline
    \multicolumn{5}{c}{\textit{Routes to CH$_3$CN$^+$}} \\
    1 &  C${_2}$H${_2^+}$ + NH$_{2}$ & $\rightarrow$ & CH$_{3}$CN$^{+}$ + H     & New QM computations  & Removed \\
    2 &  C${_2}$H$_{4}^{+}$ + N      & $\rightarrow$ & CH$_{3}$CN$^{+}$ + H     & New QM computations  & Removed \\
    3 &  NH${_2}$CHO + C$^+$       & $\rightarrow$ & CH$_{3}$CN$^{+}$ + O       & Further investigation needed & \\
    4 &  H${_2}$CCN + H${_3^+}$    & $\rightarrow$ & CH$_{3}$CN$^{+}$ + H$_{2}$ & New QM computations  & Removed\\
    \hline
    \multicolumn{5}{c}{\textit{Routes to CH$_3$CNH$^+$}} \\
    5a &  CH$_{3}$CN$^{+}$ + H$_{2}$ & $\rightarrow$ & CH$_{3}$CNH$^{+}$ + H        & Reliable & \\
    5b &  CH$_{3}$CN$^{+}$ + CH$_4$  & $\rightarrow$ & CH$_{3}$CNH$^{+}$ + CH$_3$   & Reliable & \\
    6  &  C${_2}$H${_7}{^+}$ + HCN   & $\rightarrow$ & CH$_{3}$CNH$^{+}$ + CH${_4}$ & Reliable & \\
    7a &  CH${_3}{^+}$ + HCN         & $\rightarrow$ & CH$_{3}$CNH$^{+}$ + photon   & Modified & \\
    7b &  CH${_3}{^+}$ + HNC         & $\rightarrow$ & CH$_{3}$CNH$^{+}$ + photon   & Removed & \\
    8  &  CH${_4}{^+}$ + HCN         & $\rightarrow$ & CH$_{3}$CNH$^{+}$ + H        & New QM computations & Modified\\
    9a &  CH$_{3}$OH${_2}{^+}$ + HCN & $\rightarrow$ & CH$_{3}$CNH$^{+}$ + H${_2}$O & New QM computations & Removed\\
    9b &  CH$_{3}$OH${_2}{^+}$ + HNC & $\rightarrow$ & CH$_{3}$CNH$^{+}$ + H${_2}$O & New QM computations$^*$ & Added \\
    \hline
    \multicolumn{5}{c}{\textit{Routes to CH$_3$CN}} \\ 
    10a &  CH$_{3}$CNH$^{+}$ + NH$_3$& $\rightarrow$ & CH$_{3}$CN + NH$_{4}^{+}$ & New QM computations$^*$ & Added \\
    10b &  CH$_{3}$CNH$^{+}$ + e$^-$ & $\rightarrow$ & CH$_{3}$CN + H            & New QM computations     & Modified\\
    11a &  CH$_{3}$ + CN             & $\rightarrow$ & CH$_{3}$CN + photon       & Removed  & \\
    11b &  CH$_{3}$ + CN${^-}$       & $\rightarrow$ & CH$_{3}$CN + e$^-$        & Further investigation needed & \\
    12 &  CH$_{3}$CHCH${_2}$ + CN    & $\rightarrow$ & CH$_{3}$CN + C${_2}$H$_3$ & Removed & \\
    \hline
    \end{tabular}
    \end{center}
    $^*$Reaction proposed in this work.
\end{table*}

Figure \ref{fig:overview-reactionscheme} shows the scheme of all reactions invoked in the literature and/or included in the KIDA and UDfA databases, plus two additional reactions that we propose in this work.
The reactions are identified by numbers to facilitate the following reading.
In Table \ref{fig:overview-reactionscheme}, we list the reactions and separate them in three groups, according to which precursor they form: CH$_3$CN$^+$, CH$_3$CNH$^+$ or directly CH$_3$CN, respectively. 
We notice that, the reactions leading to CH$_3$CN$^+$ are rarely taken into account in astrochemical models.

The following subsections report a critical discussion of what is known about each of the Table \ref{fig:overview-reactionscheme} reactions.
In some cases, we conclude that the database products and rate constants are not reliable and either we suggest revised values or we carry out new theoretical calculations, described in Secs. \ref{sec:methods} and \ref{sec:results}.
Table \ref{fig:overview-reactionscheme} also reports the result of our critical review, namely whether the proposed/literature reactions have rate constants and products reliable, need further investigations, have to be removed or, finally, were unreliable and we carried out new ad hoc QM computations.

\subsection{\label{subsec:overview-reaction1}
Reaction [1]: \texorpdfstring{C$_{2}$H$_{2}^{+}$ $+$ NH$_{2}$ $\rightarrow$ CH$_{3}$CN$^{+}$ $+$ H}{reaction1}}

The  KIDA and UDfA databases report four reactions forming CH$_3$CN$^+$, reactions [1] to [4] of Table \ref{tab:overview-reactionlist}. 
Reaction [1] was proposed by \citet{Woon2009}, who computed the values of polarizability and dipole moment of several species in order to derive the Langevin rate constant for many ion-neutral reactions. 
They report a BR of 50\% for the proton transfer channel that forms NH$_{3}^{+}$ + CCH, and 50\% for the channel that forms CH$_{3}$CN$^{+}$ + H.
However, reaction [1] can in principle form other products in addition to the products guessed by \citet{Woon2009}. 
Therefore, we carried out new QM computations in order to check whether CH$_{3}$CN$^{+}$  is a major product of the reaction. 
The computations results are reported in Sec. \ref{subsec:results-reaction1}.

\subsection{Reaction [2]: \texorpdfstring{C${_2}$H$_{4}^{+}$ + N  $\rightarrow$ CH$_{3}$CN$^{+}$ + H}{reaction2} \label{subsec:overview-reaction2}}

Reaction 2 is cited by \citet{loison2014interstellar} with the rate constants measured experimentally by \citet{scott1999c}. 
However, Scott et al. reported the formation of CH$_{2}$CNH$^+$ rather than CH$_3$CN$^+$, as incorrectly written in Loison et al., unfortunately without an explicit explanation why they assumed that.

To clarify the situation, we carried out the theoretical characterization of the PES to determine which products are formed.
The computations results are reported in Sec. \ref{subsec:results-reaction2}.

\subsection{Reaction [3]: \texorpdfstring{NH${_2}$CHO + C$^+ \rightarrow$ CH$_{3}$CN$^{+}$ + O}{reaction3} \label{subsec:overview-reaction3}}

As for reaction [1], \citet{Woon2009} estimated the products and the Langevin rate constants after computing the polarizability and dipole moment of the reactants.
While it would be useful to revise the products and rate constants of this reaction, since this would be very time-consuming, we decided to first assess how important this reaction is in the formation of CH$_{3}$CN$^{+}$ using astrochemical modeling (see Sec. \ref{sec:astro-modeling}).
Our conclusions after the modeling is that this route is relatively negligible.
Thus, the rate reported in the KIDA database has been used for the subsequent analysis. 

Note that, despite the reaction may be negligible in the formation of methyl cyanide, it may be important in the destruction of NH${_2}$CHO in environments where carbon is ionised.

\subsection{Reaction [4]: \texorpdfstring{H${_2}$CCN + H${_3^+} \rightarrow$ CH$_{3}$CN$^{+}$ + H$_{2}$}{reaction4} \label{subsec:overview-reaction4}}

Reaction [4] is part of the same group of the reactions [1] and [3], namely is based on the theoretical work by \citet{Woon2009} and requires revision. 
H${_2}$CCN exhibits two resonance structures, consequently it can be protonated both on the carbon atom, forming CH$_{3}$CN$^{+}$, or on the nitrogen atom, forming CH${_2}$CNH$^{+}$. 
Therefore, as for reactions [1] and [2], we carried out new QM computations in order to determine which of the two products is formed. 
The computations results are reported in Sec. \ref{subsec:results-reaction4}.


\subsection{Reactions [5]: \texorpdfstring{CH$_3$CN$^+$ + H$_{2}$ or CH$_4$ $\rightarrow$ CH$_{3}$CNH$^{+}$ + H or CH$_3$}{reaction5} \label{subsec:overview-reaction5}}

The two reactions [5a] and [5b], that form CH$_3$CNH$^+$ and then methyl cyanide via reactions [10], have been experimentally studied by \citet{martin1960hydrogen}.
The authors observed the hydrogen atom abstraction of the CH$_3$CN$^+$ ion from H$_{2}$ and CH$_4$.
In addition, they showed that this type of reactions have large rate constants (k$_{5a}$ = $5.7 \times 10^{-10}$ cm$^3$s$^{-1}$ and k$_{5b}$ = $1.7 \times 10^{-9}$ cm$^{3}$s$^{-1}$, respectively) and no appreciable temperature dependence (in the range of 100--215 $^o$C). 

We notice that reaction [5a] is present in the UDfA database but not in KIDA, while reaction [5b] is absent in both databases.
However, in KIDA, the reactions leading to the destruction of CH$_3$CN$^+$ do not form CH$_3$CN \citep{loison2014interstellar,Woon2009}. 

In summary, we include these two reactions in the network based on the validity of the experimental results.

\subsection{Reaction [6]: \texorpdfstring{C$_{2}$H$_7^+$ + HCN $\rightarrow$ CH$_{3}$CNH$^{+}$ + CH${_4}$}{reaction6} \label{subsec:overview-reaction7}}

Reaction [6] is reported in the UDfA database with a rate of production of CH$_{3}$CNH$^+$ of $10^{-10}$ cm$^3$ mol$^{-1}$ s$^{-1}$, based on \citet{mackay1980studies}.
However, only 10\% of the reactive flux actually leads to CH$_{3}$CNH$^+$. 
The channel that contributes the most is instead the one leading to the formation of HCNH$^+$ + C${_2}$H${_6}$, via a proton transfer process. 
Considering the relatively negligible abundance of C${_2}$H${_7^+}$ in the ISM, it seems unlikely that this reaction is crucial in the formation of methyl cyanide. 
For this reason, we did not perform new QM computations and adopted the rate constants reported in the work by \citealt{mackay1980studies}.

\subsection{Reactions [7]: \texorpdfstring{CH$_{3}^{+}$ + HCN or HNC $\rightarrow$ CH$_{3}$CNH$^{+}$ + photon}{reaction7} \label{subsec:overview-reaction6}}

\noindent
\textit{Reactions (7a): CH$_{3}^{+}$ + HCN $\rightarrow$ CH$_{3}$CNH$^{+}$ + photon}

The radiative association between CH$_{3}^{+}$ and HCN is believed to be the most important route to the methyl cyanide formation, followed by the dissociative recombination of CH$_{3}$CNH$^{+}$ with an electron (see also Sec. \ref{sec:astro-modeling}).
The first experimental identification of CH$_{3}$CNH$^{+}$ as product of reaction (7a) dates back to 1980 \citep{mcewan1980low}.
Later, several experiments have been performed in order to understand the mechanism of the reaction and to give a better estimate of the rate constant \citep{bass1981ion,kemper1985reexamination,knight1986isomers,anicich1986ion,anicich1993evaluated,anicich1995association,illies1981formation}. 
In the most recent one, \citet{anicich1995association} concluded that  CH$_{3}^{+}$ does react with HCN via a radiative association reaction, with a rate constant of $2 \times 10^{-10}$ cm$^{3}$s$^{-1}$ at room temperature. 
According to their experiments, the initial stabilized structure is the CH$_{3}$NCH$^{+}$ ion but, at high pressure ($> 8 \times 10^{-5}$ torr), the association complex can be efficiently stabilized through collisions, resulting in the formation of CH$_{3}$CNH$^{+}$. A similar experimental technique has been employed by \citet{luca2002low} to study the radiative association of CH$_3^+$ + H$_2$O, with promising results.

Reaction (7a) has also been object of some theoretical studies.
The rate constant at 10 K has been calculated using a statistical method and the derived parameters of the rate constant are $\alpha$ = $9 \times 10^{-9}$ and $\beta$ = -0.5 \citep{bates1983theory,herbst1985update}.
The same approach has been recently used by \citet{tennis2021radiative} to derive the rate constant for the radiative association of CH$_3$ and CH$_3$O.
However, given the large uncertainties of the method, which used several approximations, in the works by Bates and Herbst, the authors suggest that the rate constant is probably overestimated.
In a more recent work by \citet{klippenstein1996theory}, the rate constant has been estimated using a model based on a combination of ab initio QM estimates of structures, frequencies and intensities, with the VTST (Variational Transition State Theory) theory. 
The authors derived a rate constant value of $5.76 \times 10^{-9}$ cm$^{3}$s$^{-1}$ in the high efficiency limit at 300 K. 
The comparison with the trajectory simulations by \citet{su1982parametrization} shows the accuracy of the method, which only slightly overestimates the rate constant value. 
Unfortunately, the value in the low efficiency limit is not reported by the authors and the high efficiency one might be considered as an upper limit to the rate constant.

An important point to elucidate is the branching ratio of formation of CH$_{3}$CNH$^{+}$ and CH$_{3}$NCH$^{+}$, as a function of the temperature.
\citet{defrees1985theoretical} calculated the energies of the two structures CH$_{3}$CNH$^{+}$ and CH$_{3}$NCH$^{+}$ and of their isomerization barrier. 
The authors propose that as energy is lost through relaxation, whether radiatively (as it would likely occur in the ISM) or collisionally (as it occurs in experiments), the complex formed in reaction (7a) can switch between the two isomeric forms. 
This process can continue until the ion no longer has enough energy to isomerize to the other form. 
\citet{defrees1985theoretical} concluded that the ratio of formation of protonated methyl isocyanide and protonated methyl cyanide lies in the range 0.1--0.4, taking into account the uncertainty in the energy difference between the two protonated isomers. 

Later, \citet{le2019role} investigated the impact of the assumed CH$_{3}$CNH$^{+}$/CH$_{3}$NCH$^{+}$ ratio in reaction (7a) on the model predicted CH$_{3}$NC/CH$_{3}$CN abundance ratio in Photo-Dissociation Regions (PDRs).  
Comparing the astrochemical model predictions with the observed ratio, these authors concluded that the best-fit is obtained if reaction (7a) has CH$_{3}$NC/CH$_{3}$CN $\sim$ 0.2 \citep[in agreement with the theoretical calculation of][]{defrees1985theoretical}.
It has to be noted, though, that the modeling itself relied on the best fit of several other poorly known parameters, such as the elemental gaseous O/C ratio and the cosmic-ray ionization rate, so that these conclusions have to be taken with a grain of salt.

In summary, for the analysis presented in Sec. \ref{sec:astro-modeling}, we decided to use the rate constant value reported by \citet{klippenstein1996theory} and to consider that the molecule formed by the radiative association of CH$_{3}^{+}$ and HCN is CH$_{3}$CNH$^{+}$, which corresponds to 80\% of the products according to \citet{defrees1985theoretical}. 
Consequently, the formation of the isomer CH$_{3}$NCH$^{+}$, which accounts for the remaining 20\%, is not considered in the present work since it would require the inclusion of methyl isocyanide and all its formation and destruction pathways in the model, which is beyond the scope of this work.

\vspace{0.2cm}
\noindent
\textit{Reactions (7b): CH$_{3}^{+}$ + HNC $\rightarrow$ CH$_{3}$CNH$^{+}$ + photon}

The reaction of CH$_{3}^{+}$ with HNC has never been investigated, neither experimentally nor theoretically.
\citet{loison2014interstellar} assumed for this reaction the same rate constant of reaction [7a] in modelling the gas-phase reactions controlling the HCN/HNC abundance ratio.
However, this is probably a rough approximation, considering the large uncertainties on the CH$_{3}^{+}$ + HCN reaction rate constant and branching ratios. 
Moreover, while in the case of reaction [7a] there are no exothermic channels that compete with the radiative stabilization of the complex, in the case of reaction [7b] the channel of formation of CH$_{3}^{+}$ + HCN is exothermic by 14 kcal mol$^{-1}$. 
The presence of this exothermic channel identified by \citet{defrees1985theoretical}, makes the radiative association process unlikely with respect to the normal ion-molecule reaction. 
Therefore, we decided to remove this reaction from our network.

\subsection{Reaction [8]: \texorpdfstring{CH$_{4}^{+}$ + HCN $\rightarrow$ CH$_{3}$CNH$^{+}$ + H}{reaction8} \label{subsec:overview-reaction8}}

Reaction [8] has been experimentally studied by \cite{mcewan1981icr} and \cite{anicich1986ion}. 
They measured a rate constant of $3.3\times10^{-9}$ cm$^{3}$s$^{-1}$ at room temperature, for the proton transfer channel that forms HCNH$^+$ + CH$_3$. 
A competitive channel forming C$_2$H$_4$N$^+$ + H has also been identified and reported with a BR of 0.02. 

We notice that KIDA incorrectly reports the rate constant of formation of CH$_{3}$CNH$^{+}$ without taking into account the actual formed small fraction, $\leq$2\%. 
Moreover, although \citet{mcewan1981icr} and \citet{anicich1986ion} do not explicitly quote it, the reaction probably forms CH$_{3}$NCH$^+$ rather than CH$_{3}$CNH$^+$, when considering that the nitrogen atom attacks the CH$_{4}^{+}$ ion.

A theoretical study of reaction [8] of relevance to Titan’s ion chemistry was already proposed by \citet{li2008theoretical}. 
They report the PES of the reaction, with the geometries calculated at B3LYP/6-311G(d,p) level and then the single point energies of the optimised geometries calculated at CCSD(T)/6-311G++(3df,2pd) level. 
However, we decided to investigate the reaction between HCN and CH$_{4}^{+}$ with an higher level of theory, focusing only on the exothermic channels of astrochemical interest. 
We also performed kinetic calculations to compute the rate constant of formation of each product.
The results are reported in Sec. \ref{subsec:results-reaction8}

\subsection{Reaction [9]: \texorpdfstring{CH$_3$OH$_{2}^{+}$ + HCN or HNC $\rightarrow$ CH$_{3}$CNH$^{+}$ + H${_2}$O}{reaction9} \label{subsec:overview-reaction9}}

\noindent
\textit{Reactions [9a]: CH$_3$OH$_{2}^{+}$ + HCN $\rightarrow$ CH$_{3}$CNH$^{+}$ + H${_2}$O}

Reaction [9a] is reported in UDfA as a possible source of CH$_{3}$CNH$^+$. 
The reaction has been investigated experimentally by \citet{meot1986ion}, who measured a rate constant of $2.3\times10^{-11}$ cm$^3$ s$^{-1}$ at 340 K. 
However, this reaction should rather form CH$_{3}$NCH$^+$ since the electron-rich nitrogen atom would preferably attack the CH$_{3}$ group and form protonated methyl isocyanide (CH$_{3}$NCH$^+$) instead. 
This is suggested also by the authors of the experimental study, based on proton affinity considerations \citep{meot1986ion}.
As in other cases, no theoretical study is available in the literature for this reaction. 

The role of protonated methanol in the synthesis of iCOMs was first suggested by \cite{taquet2016formation} and successively  discussed by \citet{skouteris2019interstellar} in the framework of the formation of dimethyl ether.
In regions where methanol is very abundant, such as hot cores, hot corinos and molecular shocks, this route of formation of methyl cyanide could be important.
For this reason, in this work, we carried out new QM computations to verify whether the product is CH$_{3}$NCH$^+$ or CH$_{3}$CNH$^+$. 
The results of the computations are reported in Sec. \ref{subsec:results-reaction9}.

\noindent
\textit{Reactions [9b]: CH$_3$OH$_{2}^{+}$ + HNC $\rightarrow$ CH$_{3}$CNH$^{+}$ + H${_2}$O}

Since the reaction of protonated methanol with HCN likely produces CH$_{3}$NCH$^{+}$ rather than CH$_{3}$CNH$^{+}$, we also performed new QM computations of the reaction [9b].
The results of the computations are reported in Sec. \ref{subsec:results-reaction9}.

\subsection{Reactions [10]: \texorpdfstring{CH$_3$CNH$^+$ + NH$_3$ or e$^-$ $\rightarrow$ CH$_{3}$CN + NH$_{4}^{+}$ or H}{reaction10} \label{subsec:overview-reaction10}}

\noindent
\textit{Reaction [10a]: CH$_3$CNH$^+$ + NH$_3$ $\rightarrow$ CH$_{3}$CN + NH$_{4}^{+}$} \label{subsec:protontrasferammonia}

The proton transfer to ammonia, firstly included in astrochemical models by \citet{rodgers2001chemical}, is an alternative process to the electron recombination, that converts the molecular ion, in our case CH$_{3}$CNH$^+$, into its neutral counterpart, CH$_{3}$CN. 
The rates of several proton transfer processes are included in (some) astrochemical models but the parameters are often guessed considering similar reactions. 
For instance, \citet{taquet2016formation} assumed that all the proton transfer reactions are non-dissociative and occur with a rate of $4\times 10^{-9}$ cm$^{3}$s$^{-1}$ at 100 K.
This assumption is based on the work of \citet{hemsworth1974rate} at room temperature, in which the rate constants for the proton transfer reactions to ammonia are all reported having the same order of magnitude, $2\times 10^{-9}$ cm$^{3}$s$^{-1}$. 
However, relatively small differences (up to a factor 10) in these rate constants could change the results of the astrochemical model predictions, given the high non linearity of the equations.
For example, QM computations by \citet{skouteris2019interstellar} of the (CH$_3$)$_2$OH$^+$ + NH$_3$ gave a rate constant equal to $1\times 10^{-9}$ cm$^{3}$s$^{-1}$ in the 10--300 K range. 

For this reason, we carried out new QM computations of this reaction. 
The results are reported in Sec. \ref{subsec:results-reaction10}.

\vspace{0.2cm}
\noindent
\textit{Reactions [10b]: CH$_3$CNH$^+$ + e$^-$ $\rightarrow$ CH$_{3}$CN + H}\label{subsec:electronrecombination}


Reaction [10b] is a particular case of electron-ion recombination reaction. It is normally accompanied by the dissociation of the neutralized species since the large amount of energy released is enough to break one or more bonds \citep{geppert2006dissociative}. While in many cases there are multiple bond fissions and the fission of C-C bonds is expected, in the case of the CH$_3$CNH$^+$ + e$^-$ reaction a large fraction of the products preserve the C-C-N skeleton, and one or more H atoms are instead lost \citep{geppert2007formation}. Indeed, this reaction has been object of several experimental studies. In the work by \citet{vigren2008dissociative} and \citet{geppert2007formation}, the authors derived a thermal rate constant for the electron recombination of CD$_3$CND$^+$ of $8.1 \times 10^{-7} (\frac{T}{300})^{-0.69}$ cm$^{3}$s$^{-1}$ using the CRYRING ion storage ring (SR). They also report the presence of two main dissociative recombination channels: the ones preserving the carbon-chain (65\%) and those breaking one bond between heavy atoms and leading to more fragmented species (such as HCN, HNC and CH$_{3}$) (35\%). Unfortunately, the authors did not report the specific products of the CH$_3$CNH$^+$ + e$^-$ reaction. In the work by \citet{mclain2009flowing} instead, flowing afterglow experiments have been used to measure the rate constant of both the monomer CH$_3$CNH$^+$ and the dimer (CH$_{3}$CN)$_2$H$^+$. In the first case they found a power law dependence of $T^{-1.03}$ while in the second case of $T^{-0.5}$. 

Since the SR measurement was made with fully deuterated ions and there is an evidence that the rate constants can differ significantly when different isotopes are used \citep{guberman2003dissociative}, we decided to use the value reported by \citet{mclain2009flowing} for the monomer (e.g. $3.4 \times 10^{-7} (\frac{T}{300})^{-1.03}$ cm$^{3}$s$^{-1}$) in the further analysis. 

Regarding the specific formation of methyl cyanide, \citet{loison2014interstellar} used branching fractions (BF) and rate constants of four different channels guessed from the work of \citet{plessis2012production}.
However, the products could be different from the reported ones. 
Indeed, there are several pathways of fragmentation and retention of carbon chain which can occur and the description of this system may be more complicated than the one cited by \citet{loison2014interstellar}. 
In this paper, starting from the experimental works of \citet{geppert2007formation} and \citet{mclain2009flowing}, we improved the work by \citet{loison2014interstellar} focusing on the BFs of the products in which the carbon chain is maintained, with the goal of determine the exact BF of formation of CH$_3$CN. The computations results are reported in Sec. \ref{subsec:results-reaction10}.

\subsection{Reactions [11]: \texorpdfstring{CH$_3$ + CN or CN$^-$ $\rightarrow$ CH$_{3}$CN + photon or e$^-$}{reaction11} \label{subsec:overview-reaction11}}

Reactions [11a] and [11b] can form CH$_3$CN directly. 
The first one is a radiative association while the second one is an associative detachment. 
These two processes have been proposed by \citet{prasad1980}, but the rate constants of both reactions are only guessed and have not been studied via a specific theoretical or experimental work. 

Reaction [11a] is reported by \citet{loison2014interstellar} with an alternative channel that forms H$_{2}$CCN + H. 
This is based on the work by \citet{yang2005global}, in which the PES of the CH$_3$CN decomposition is computed. 
In particular, the reaction between CH$_3$ and CN produces the CH$_3$CN complex, which can then dissociate into CH$_2$ + HCN with an energy barrier that is below the reactants' energy, albeit the H$_2$CCN + H products are not reported. 
Since the presence of a competitive exothermic channel makes the radiative association path unlikely, we decided to remove reaction [11a] from our network.


We also decided not to revise reaction [11b] since, based on the astrochemical modeling of Sec. \ref{sec:astro-modeling}, its role in the formation of CH$_3$CN is not relevant.

\subsection{Reaction [12]: \texorpdfstring{CH$_3$CHCH$_{2}$ + CN $\rightarrow$ CH$_{3}$CN + C${_2}$H$_3$}{reaction12} \label{subsec:overview-reaction12}}

Another reaction reported in the KIDA and UDfA databases for the direct formation of methyl cyanide is that between propylene (CH$_3$CHCH$_{2}$) and the CN radical.
This reaction, however, is not correct. 
In fact, \citet{sims1993rate} measured the total rate constant as a function of the temperature, using the CRESU technique, but could not determine the products of the reaction.
A later theoretical study by \citet{gannon2007h} showed that CH$_3$CN is not a product of the reaction. 
We, therefore, removed this reaction from the network.

\subsection{Summary} \label{subsec:overview-summary}

We have reviewed a total of seventeen reactions eventually leading to the formation of methyl cyanide and that are present in the literature and/or in the astrochemical databases KIDA and UDfA.
They can be separated into three groups, forming CH$_3$CN$^+$, CH$_3$CNH$^+$ and finally CH$_3$CN, respectively, as detailed in the following.

\vspace{0.2cm}
\noindent
\textit{CH$_3$CN$^+$ formation}:
A first group of four reactions form CH$_3$CN$^+$, which then forms CH$_3$CNH$^+$ and, finally, CH$_3$CN. 
We decided to carry out new QM computations of the three most important channels for the CH$_3$CN$^+$ formation, the reactions [1], [2] and [4] of Tab. \ref{tab:overview-reactionlist}.
Based on the astrochemical modeling of Sec. \ref{sec:astro-modeling}, reaction [3] has a negligible role in the methyl cyanide formation, so we leave it to further investigations.

\vspace{0.2cm}
\noindent
\textit{CH$_3$CNH$^+$ formation}:
Eight reactions lead to the CH$_3$CNH$^+$ formation.
Three of them, reactions [5a], [5b] and [6], have been previously studied and are correctly reported in the astrochemical KIDA and UDfA databases.
We carried out new QM computations for the newly proposed reaction [9b] and for reactions [8] and [9a], which needed revision.
Reaction [7a] is used with an updated rate constant value derived from the literature, while reaction [7b] is removed from the network due to the presence of an exothermic channel.

\vspace{0.2cm}
\noindent
\textit{CH$_3$CN formation}:
Five reactions lead to the CH$_3$CN formation.
We carried out new QM computations for the newly proposed reaction [10a] and for reaction [10b], which needed revision. 
Reactions [11a] is removed since it has an exothermic channel as reaction [7b], while reaction [11b] would need further investigation, not done in this work because likely unimportant.
Finally, we removed reaction [12] since it has been studied and incorrectly included in the astrochemical KIDA and UDfA databases.

\vspace{0.2cm}
\noindent
Table \ref{tab:overview-reactionlist} summarizes the results of the critical review.

\section{Computational Methods}\label{sec:methods}

Based on the review of the published articles and our astrochemical simulations, discussed in the previous section (\S ~\ref{sec:overview-gas-routes}), we decided to carry out new QM computations of eight reactions.
To this end, we first computed the Potential Energy Surface (PES) of each system and identified the products, as described in Sec. \ref{subsec:methods-pes}.
We then computed the rate constants of each channel, i.e. the capture rate constant of each reaction and the branching fractions using the Rice-Ramsperger-Kassel-Markus (RRKM) theory, as described in Sec. \ref{subsec:methods-kinetics}.
The theory makes use of the standard functional $V(r)$ for the potential energy of the two approaching species.
In one case, we found that the determination of the so-called $C_4$ coefficient needed a specific study, and we used a semi-empirical model described in Sec. \ref{subsec:methods-semiemp}.

\subsection{Electronic structure calculations} \label{subsec:methods-pes}

The PES of the studied systems has been characterized by optimizing the various stationary points. 
We used the same computational procedure described in previous works of our group  \citep{rosi2013theoretical,skouteris2019interstellar}, in which the geometries of minima and saddle points have been first optimized with a cost-effective method.
Subsequently, single-point calculations were performed employing an expensive method to obtain accurate energies. 

For the first step, all the calculations were performed according to the Density Functional Theory (DFT) with the hybrid B3LYP \citep{becke1993density,stephens1994ab} functional in conjunction with the correlation consistent valence polarized basis set aug-cc-pVTZ \citep{dunning1989gaussian}. 
The nature of each stationary point, identified by geometry optimization, has been determined by performing harmonic vibrational frequencies calculations. 
Moreover, the nature of the saddle points has been assigned through Intrinsic Reaction Coordinate (IRC) calculations \citep{gonzalez1990reaction,gonzalez1989improved}. 
In order to obtain more accurate values of the energy of each stationary point, coupled-cluster single and double excitations augmented by a perturbative treatment of the triple excitations (CCSD(T)) \citep{bartlett1981many,raghavachari1989fifth,olsen1996full} calculations have been performed with the same basis set.
The zero-point-energy (ZPE) correction, computed using the harmonic vibrational frequencies obtained at the B3LYP/aug-cc-pVTZ level of theory, has been added to the computed CCSD(T) energies in order to correct them at 0 K. 

All calculations have been performed with the GAUSSIAN 09 \citep{g09} and GAUSSIAN 16 \citep{g16} packages, while the analysis of the vibrational frequencies was performed using the Avogadro package \citep{hanwell2012avogadro}. 

\subsection{Kinetic calculations} \label{subsec:methods-kinetics}

Kinetic calculations were performed using a combination of capture and RRKM theories, as described in previous works of our group \citep{rosi2013theoretical,skouteris2019interstellar}.
The initial bimolecular rate constant leading from the reactants to the first intermediate has been evaluated using capture theory calculations.
Capture theory accounts for the long-range attractive forces and assumes that the rate-determining step of a reaction is the formation of an intermediate reaction complex, after which products are formed with unit probability. 
The energy values for the approaching reactants are fitted to a functional which is: 
\begin{equation}
    V(r)= -\frac{C_n}{r^n} + b
	\label{eq:capture}
\end{equation}
where r is the distance between the two species, C$_n$ is the proportionality constant between the potential and $1/r^n$, usually determined by fitting the long-range ab initio data, and $n$ is a parameter that depends on type of the system (6 for neutral--neutral and 4 for ion--neutral pairs). Moreover, $b$ is an empirical factor that, for each selected configuration of the reagents,  corrects the radial dependence of the interaction for contributions that at long range add to dispersion and induction attraction components.
The capture coefficient is used to evaluate the capture cross section with the formula:
\begin{equation}
    \sigma(E)= \pi \times 2 \times \left({\frac{C_4}{E}}\right)^{1/2}
	\label{eq:crosssection}
\end{equation}
where E is the translational energy. 
The cross section is then multiplied by the collision velocity $(2E/\mu)^{1/2}$, where $\mu$ is the reduced mass of the system, to obtain the capture rate constant. 

Once the intermediate is formed, it can dissociates back to reactants, and the unimolecular rate constant for this back-dissociation is given by the equation:
\begin{equation}
    k_{back}= k_{capt} \times \frac{\rho(R)}{\rho(I)}
	\label{eq:backdissociation}
\end{equation}
where $k_{capt}$ is the capture rate constant, $\rho$(R) is the density of states per unit volume for the reactants and $\rho$(I) is the density of states for the intermediate. 

Finally, we have performed RRKM calculations as follows. 
The microcanonical rate constant is calculated using the formula:
\begin{equation}
    k_{E}= \frac{N(E)}{h\rho(E)}
\end{equation}
where N(E) denotes the sum of states in the transition state at energy E, $\rho$(E) is the reactant density of states at energy E and h is Planck's constant. 


Knowing the capture rate constant and the unimolecular rate constants, it is possible to solve the master equation for the system and consequently derive the branching ratios for the overall reaction as a function of the energy. 
After that, the rate constants for the formation of each product are determined multiplying the branching ratio for the capture rate constant. 
Finally, Boltzmann averaging was carried out to obtain temperature-dependent rate constants.
The new rate constants were then fitted with the form used in the astrochemical databases and models:
\begin{equation} \label{eq:Arrhenius}
   k(T)=\alpha \left(\frac{T}{300 K}\right)^\beta ~exp\left(-\frac{\gamma}{T}\right)
\end{equation}

\subsection{Semi-empirical model} \label{subsec:methods-semiemp}

In some cases, specifically for the reactions [9a] and [9b], we used a different method to derive the coefficient $C_4$, introduced in the previous Section, for the reasons that will be explained in Sec. \ref{subsec:results-reaction9}.
This was obtained by using a semi-empirical potential function described in the following and already used by our group \citep{valencca2022semiempirical}.

At large and intermediate separation distances, the total interaction potential $V_{total}$ is assumed to be:
\begin{equation}
    V_{total} = V_{ILJ}+V_{electr}
	\label{eq:vtot}
\end{equation}
where $V_{electr}$ is the electrostatic interaction contribution and $V_{ILJ}$ is the contribution due to non-electrostatic interaction, namely size repulsion plus induction and dispersion attraction. 
The electrostatic contribution has been evaluated considering the Coulombic interactions between pairs of ESP charges obtained from the CCSD(T)/aug-cc-pVTZ calculations for each atom of the two interacting fragments: 
\begin{equation}
    V_{electr} = \frac{1}{4\pi\epsilon_0}\sum_{i}\sum_{j}\frac{q_i q_j}{r_{ij}}
	\label{eq:velectr}
\end{equation}
The non-electrostatic term $V_{ILJ}$ has been evaluated using the so called Improved Lennard-Jones (ILJ) potential function, which provides a better reproduction with respect to the classical Lennard-Jones model over both short- and long-range distances \citep{pirani2008beyond,cappelletti2008bond}. 
The ILJ potential can be expressed as: 

\
\begin{equation}
    V_{ILJ} = \epsilon\left[\frac{m}{n(r)-m}\left(\frac{r_m}{r}\right)^{n(r)} - \frac{n(r)}{n(r)-m}\left(\frac{r_m}{r}\right)^{m}\right]
    \label{eq:vilj}
\end{equation}
where r is the distance between the center of mass of the reactants, $\epsilon$ and $r_m$ are the well depth and position of the potential, respectively, $m=4$ for ion-neutral systems and $n(r)$ can be expressed as follows: 
\begin{equation}
    n(r) = \beta + 4 \left(\frac{r}{r_m}\right)^2
    \label{eq:n(r)}
\end{equation}
%

$\beta$ is a constant parameter within wide classes of systems, and depends on the nature and hardness of the interacting particles. In our case it is set to 8 since this value is suggested for those cases where the long-range attraction is dominated by electrostatic forces like the ion-permanent dipole and ion–ion components.
The $r_m$ and $\epsilon$ parameters, of relevance for the present study, have been evaluated from  polarizability of the two interacting partners obtained from ab initio calculations.

\section{Results of the QM computations} \label{sec:results}

In this section we report the results of the PES and kinetics calculations for four studied reactions ([9a], [9b], [10a] and [10b]). Three of these reactions ([9b], [10a] and [10b]) are particularly important for the formation of CH$_3$CN according to the modeling described in Sec. \ref{sec:astro-modeling}, while reaction [9a] is a case study for which the semi-empirical model described in Sec. \ref{subsec:methods-semiemp} has been used.  The results for the other four studied reactions ([1], [2], [4] and [8]) are reported in Sec.\ref{sec:appendix-results}, being less relevant for the formation of CH$_3$CN .

\subsection{Reaction [1]: \texorpdfstring{C${_2}$H${_2^+}$ + NH$_{2}$ $\rightarrow$ CH$_{3}$CN$^{+}$ + H}{reaction1} \label{subsec:results-reaction1}}

We carried out new QM computations, whose details are reported in \ref{APP-subsec:results-reaction1}, and found that the CH$_{3}$CN$^{+}$ is a very minor product, as it is much less exothermic than other products (by more than 200 kJ mol$^{-1}$) and it presents high energy barriers along the reaction pathway.
We, therefore, decided to remove this reaction from the network.

\subsection{Reaction [2]: \texorpdfstring{C${_2}$H$_{4}^{+}$ + N  $\rightarrow$ CH$_{3}$CN$^{+}$ + H}{reaction2} \label{subsec:results-reaction2}}

Our QM computations, whose details are reported in \ref{APP-subsec:results-reaction2}, show that CH$_2$NCH$^+$, and not CH$_{3}$CN$^{+}$, is the major product of the reaction.
We, therefore, removed this reaction from the network.

\subsection{Reaction [4]: \texorpdfstring{H${_2}$CCN + H${_3^+}$ $\rightarrow$ CH$_{3}$CN$^{+}$ + H$_{2}$}{reaction4} \label{subsec:results-reaction4}}

Our QM computations, whose details are reported in \ref{APP-subsec:results-reaction4}, show that CH$_2$NCH$^+$, and not CH$_{3}$CN$^{+}$, is the major product of the reaction.
We, therefore, removed this reaction from the network as it does not contribute to the CH$_3$CN formation.

\subsection{Reaction [8]: \texorpdfstring{CH$_{4}^{+}$ + HCN $\rightarrow$ CH$_{3}$CNH$^{+}$ + H}{reaction8} \label{subsec:results-reaction8}}

Our QM computations, whose details are reported in \ref{APP-subsec:results-reaction8}, show that CH$_3$ + HCNH$^+$ are the major products of the reaction: CH$_3$CNH$^+$ is only 1\%.

\subsection{Reactions [9]: \texorpdfstring{CH$_3$OH$_{2}^{+}$ + HCN or HNC $\rightarrow$ CH$_{3}$CNH$^{+}$ + H${_2}$O}{reaction9} \label{subsec:results-reaction9}}

\begin{figure*}
	\includegraphics[width=8.8cm]{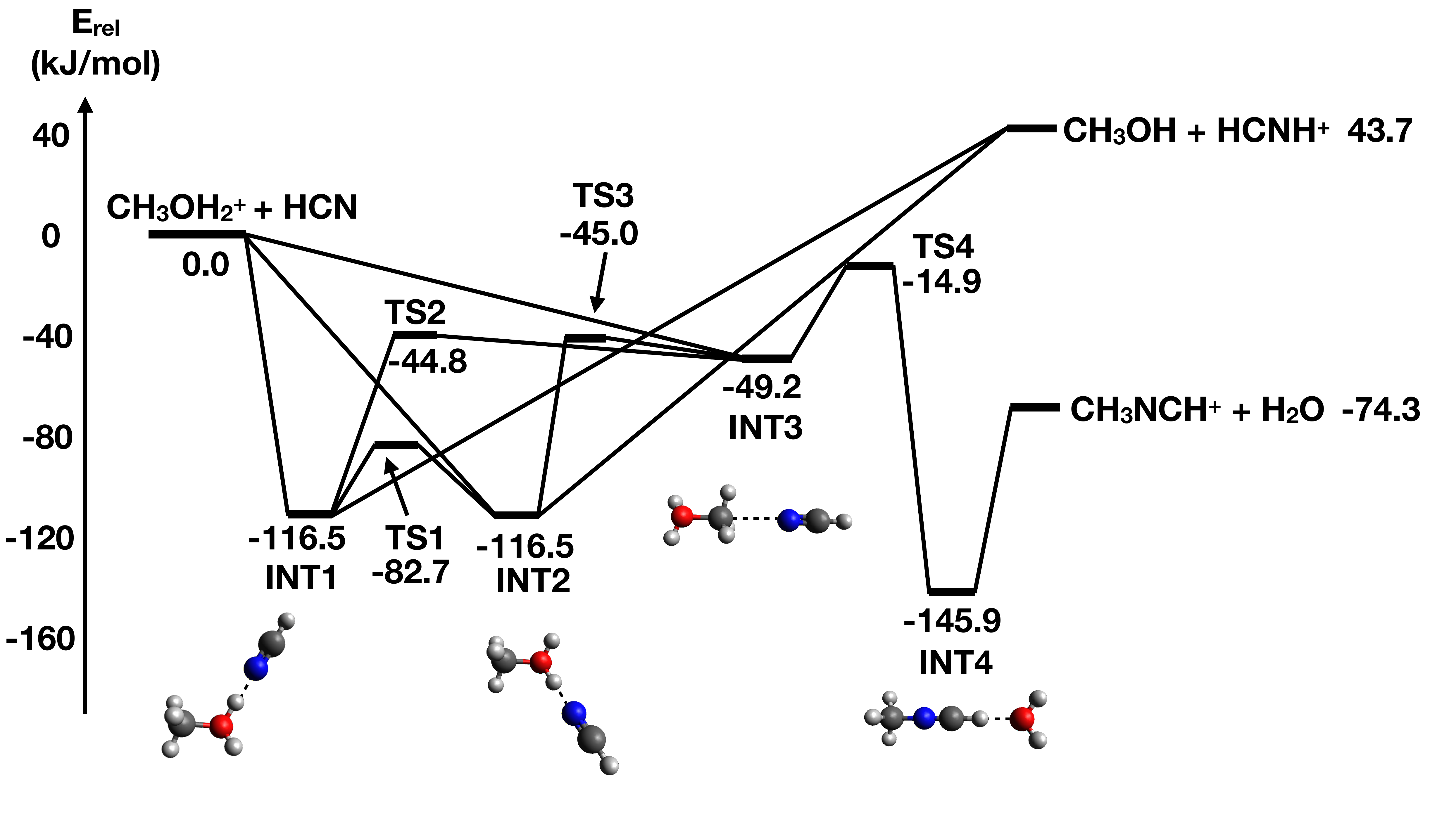}
	\includegraphics[width=8.8cm]{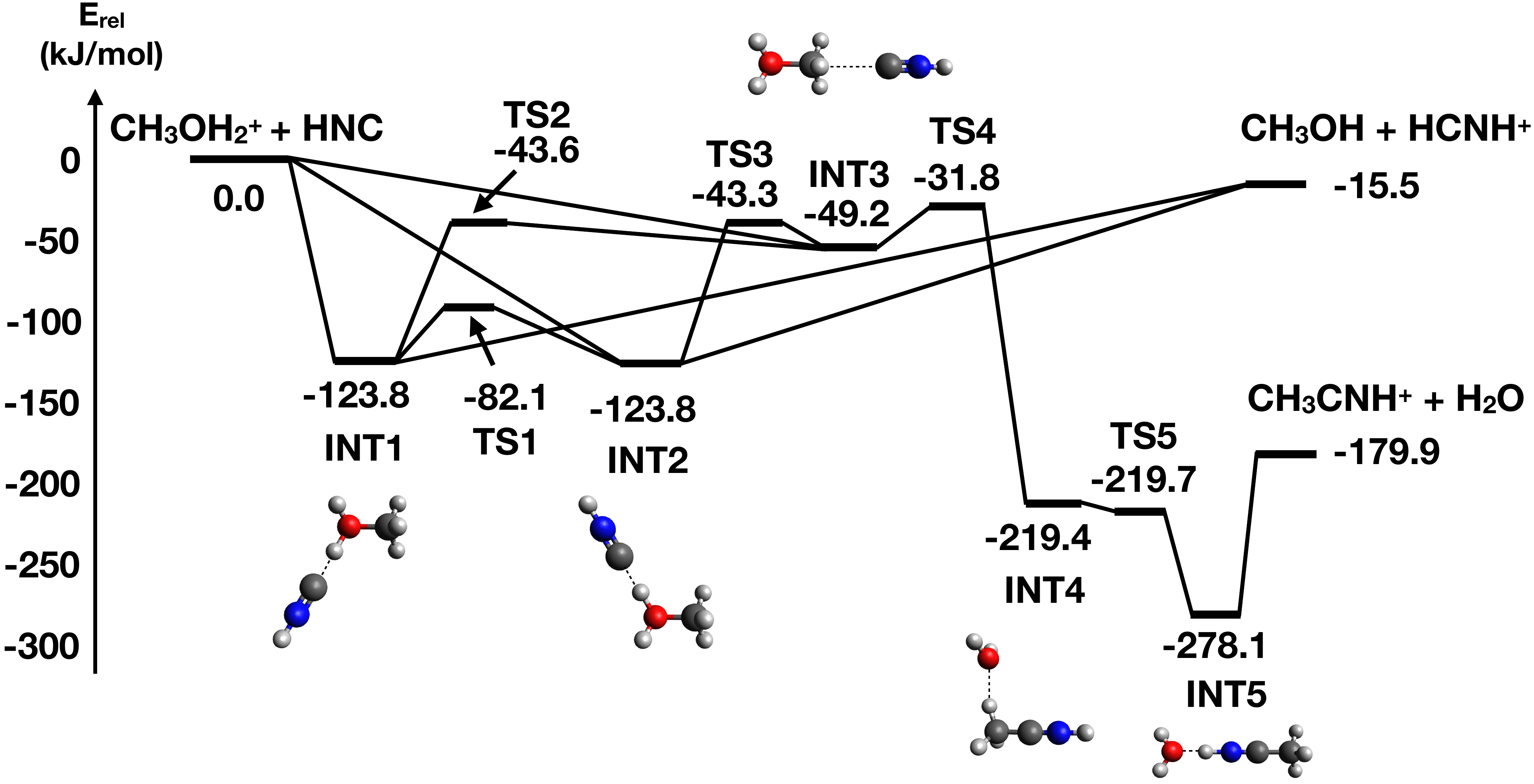}
    \caption{PES of the reactions [9a] (\textit{left panel}) and [9b] (\textit{right panel}) of Table \ref{tab:overview-reactionlist}.
    The QM computations are obtained at the CCSD(T)/aug-cc-pVTZ//B3LYP/aug-cc-pVTZ level of theory. }
    \label{fig:pes-reac9}
\end{figure*}

Since computations on reactions [9a] and [9b] have never been published and these two reactions are potentially important in the formation of methyl cyanide, we discuss here in detail their cases.
We anticipate the results, for the reader not interested in the details and that prefers skip them.
Reaction [9a] does not form methyl cyanide but rather its isomer, methyl isocyanide.
Therefore, we removed this reaction.
On the contrary, reaction [9b], never previously considered in the literature, forms methyl cyanide and hence, was added to the final network. 

\subsubsection{PES of reactions [9]}

\noindent
\textit{Reaction [9a]: CH$_3$OH$_{2}^{+}$ + HCN $\rightarrow$ CH$_{3}$NCH$^{+}$ + H${_2}$O}

Figure \ref{fig:pes-reac9} (left panel) reports the PES of reaction [9a].
From the reactants, three intermediates can be formed: INT1 and INT2, in which the nitrogen atom points toward one of the two hydrogen atoms in the OH$_{2}$ group, or INT3, in which the nitrogen atom points the carbon atom of the CH$_3$ group. 

INT1 can isomerize to INT2 and INT2 to INT1, through a barrier of 33.8 kJ mol$^{-1}$, represented by TS1.
Alternatively, the HCN molecule can move towards the CH$_3$ group and form INT3, overcoming a barrier of 71.7 kJ mol$^{-1}$ and 71.5 kJ mol$^{-1}$ (TS2 and TS3, respectively). 
INT1 and INT2 can also form CH$_3$OH and HCNH$^+$ via a proton transfer process. 
Even if no barrier is involved in the process, this channel is endothermic by 43.7 kJ mol$^{-1}$. 

When INT3 is formed, it can isomerize to INT4 overcoming a barrier of 34.3 kJ mol$^{-1}$ (TS4), in which a new C-N bond is formed while the C-O bond is broken. 
This allows the water molecule to separate from the intermediate and form the CH$_3$NCH$^+$ + H$_{2}$O products, localised at -74.3 kJ mol$^{-1}$ with respect to the reactants. 
Therefore, the only exothermic channel of the reaction is the one that forms methyl isocyanide, in agreement with what suggested by \citet{meot1986ion}.

\vspace{0.3cm}
\noindent
\textit{Reaction [9b]: CH$_3$OH$_{2}^{+}$ + HNC $\rightarrow$ CH$_{3}$CNH$^{+}$ + H${_2}$O}

Figure \ref{fig:pes-reac9} (right panel) reports the PES of reaction [9b]. 
The case of the reaction of protonated methanol with HNC is basically the same as with HCN, with two major differences.
First, the proton transfer channel is exothermic by 15.5 kJ mol$^{-1}$, making the formation of CH$_3$OH + HCNH$^+$ a competitive process to the formation of the product of interest.
Second, TS4 does not form directly the linear intermediate INT5, but instead it forms first a perpendicular intermediate INT4, and then INT5 through TS5 which is localised as a submerged transition state. 

\vspace{0.3cm}
Two characteristics can be noticed in the reactions [9a] and [9b]. 
The first one is that, while the proton transfer is a direct process which does not involve any transition state to be overcome, the formation of CH$_3$CNH$^+$ requires at least one or two isomerizations of the intermediates. 
The second characteristics is that there is an inverse relation between the reactant and the product formed: if the reactant is HCN only the isocyano product can be formed (CH$_3$NCH$^+$), while if the reactant is HNC only the cyano product is formed (CH$_3$CNH$^+$). 

\subsubsection{Kinetics of reactions [9]}

\noindent
\textit{Reaction [9a]: CH$_3$OH$_{2}^{+}$ + HCN $\rightarrow$ CH$_{3}$CNH$^{+}$ + H${_2}$O}

Although the CH$_3$OH$_{2}^{+}$ + HCN reaction does not form protonated methyl cyanide, we carried out the kinetic calculations because it has been experimentally studied by \citet{meot1986ion}, which gives us a very important constraint on the value of its rate constant, $2.3 \times 10^{-11}$ cm$^3$ s$^{-1}$ at 340 K.
We, therefore, studied this reaction with the aim to test our computations against the experiments and, then, apply what learned to reaction [9b] which does form protonated methyl cyanide. 


As a first attempt, we derived the capture coefficient from the fit of the ab initio data and we obtained a C$_4$ coefficient equal to 12.7 Ha \r{A}$^4$ and a capture rate constant of $4.2 \times 10^{-9}$ cm$^3$ s$^{-1}$. 
We then performed the full kinetic calculations and we obtained a rate constant for the formation of CH$_{3}$CNH$^{+}$ + H${_2}$O at 340 K of $7.2\times10^{-10}$ cm$^3$ s$^{-1}$, which is more than one order of magnitude higher than the experimental value found by \citet{meot1986ion}.

In order to obtain a value in better agreement with the experimental one, we modeled the entrance potential with a more accurate method using the semi-empirical potential described in Sec. \ref{subsec:methods-semiemp}.
Specifically, the C$_4$ coefficient was obtained as follows. 
We first computed it by averaging all over the possible configurations of the reactants and assuming that all the orientations are reactive. 
Therefore, by definition, the electrostatic contribution is zero because of the compensation between attractive and repulsive effects. 
With this method we obtained an average C$_4$ coefficient of 0.78 Ha \r{A}$^4$ and a rate constant of formation of CH$_{3}$CNH$^{+}$ + H${_2}$O that was still one order of magnitude too high ($1.8\times10^{-10}$ cm$^3$ s$^{-1}$ at 340 K).

\begin{figure*}
	\includegraphics[width=8.5cm]{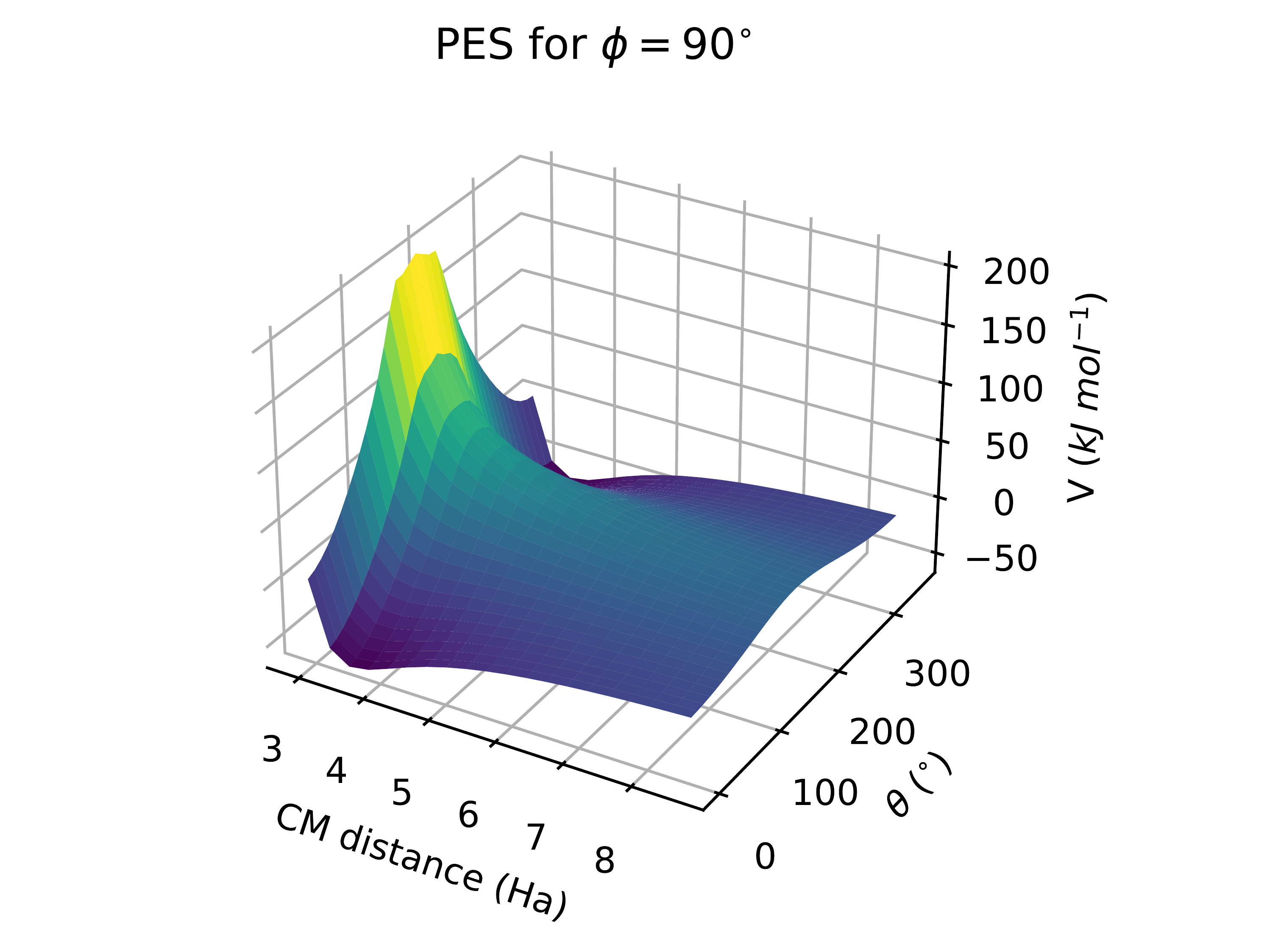}
	\includegraphics[width=8.8cm]{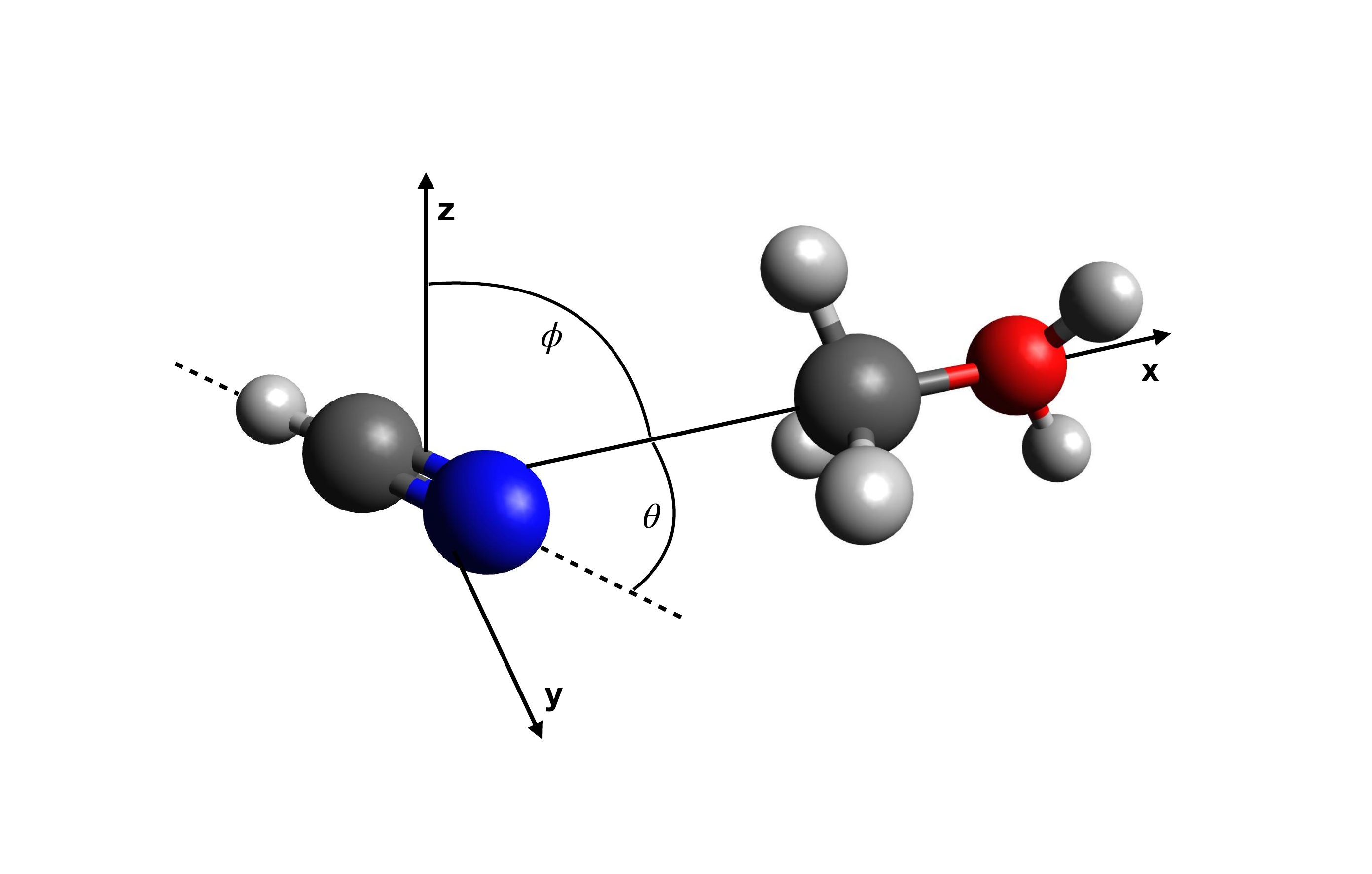}
    \caption{\textit{Left panel:} Calculated entrance potentials of reaction [9a], CH$_3$OH$_{2}^{+}$ + HCN, as a function of the rotation of HCN around its center of mass, using the semi-empirical model described in Sec. \ref{subsec:methods-semiemp}. 
    \textit{Right panel:} The plot shows the adopted reference system and coordinates used for the representation of the interaction potential between HCN and CH$_3$OH$_2^+$. 
    The rotation of HCN is obtained by keeping the $\phi$ angle fixed (90$^{\circ}$) while the $\theta$ angle varies. 
    For each combination of $\phi$ and $\theta$ the HCN molecule is translated varying the distance between the centre of mass of the two molecules and the potential is evaluated at each step.
    }
    \label{fig:results-pesreac9rot}
\end{figure*}

We, therefore, decided to evaluate the electrostatic contribution to the total entrance potential for each specific configurations of the reactants. 
We found that all the orientations with the nitrogen atom of HCN pointing CH$_3$OH$_{2}^{+}$ are attractive, while all the orientations with the hydrogen atom of HCN pointing CH$_3$OH$_{2}^{+}$ are repulsive (see Fig.\ref{fig:results-pesreac9rot}). 
We then concluded that this reaction shows an important steric effect which leads to a lower rate constant. 
In order to evaluate the steric effect, we developed a code which rotates the HCN molecule around its center of mass and calculates the potential energy curve for each configuration of the reactants. 
An example of the obtained potentials is shown in Fig. \ref{fig:results-pesreac9rot}.




%
\begin{figure*}
    \includegraphics[width=8.2cm]{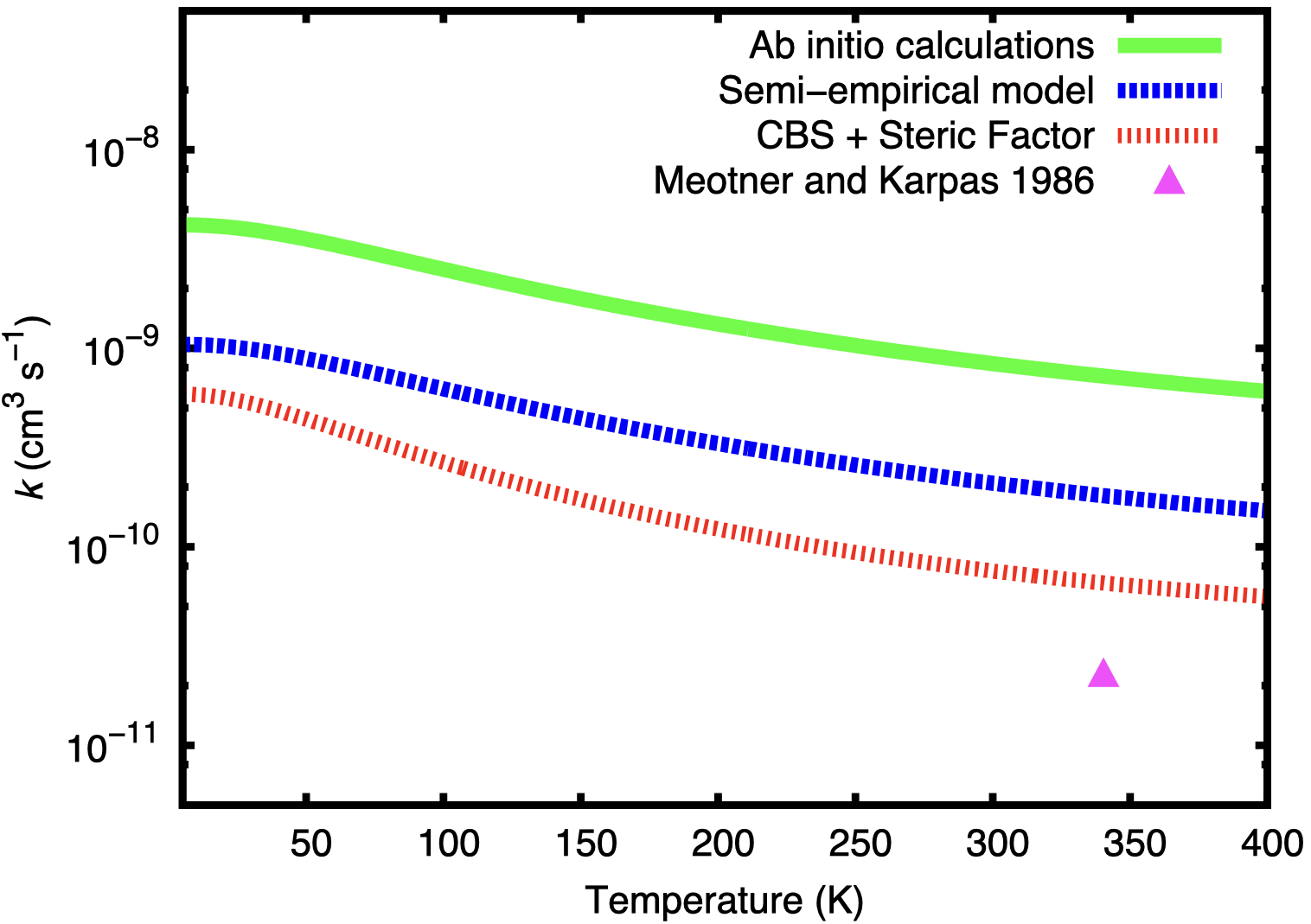}
    \includegraphics[width=8.2cm]{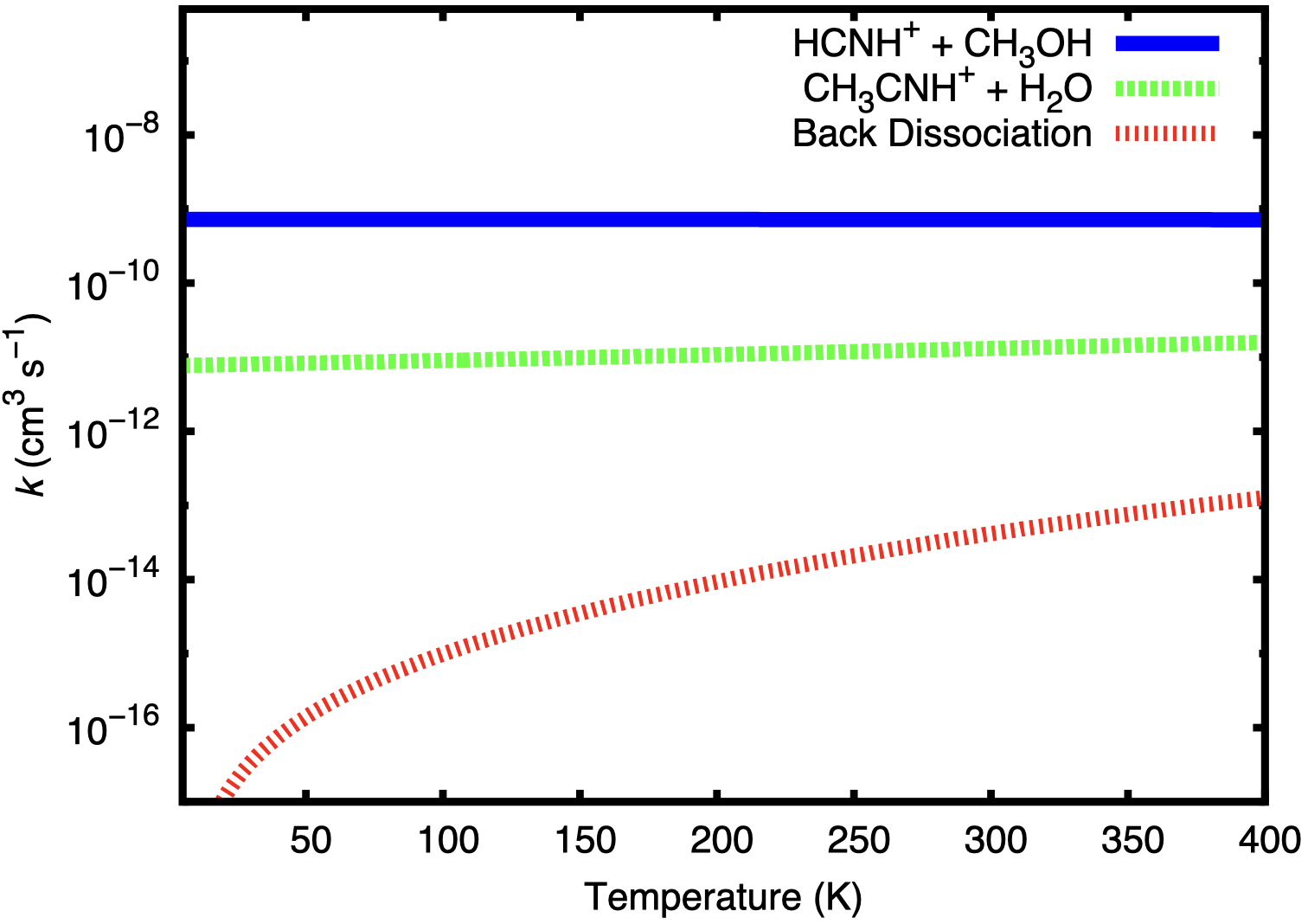}
    \caption{\textit{Left panel:} Rate constants of reaction [9a], CH$_3$OH$_{2}^{+}$ + HCN, as a function of the temperature, for the formation of the products CH$_3$NCH$^+$ + H$_{2}$O.
    The green curve shows the rate constant first obtained from ab initio computations.
    The blue curve shows the values when the semi-empirical method, described in Sec. \ref{subsec:methods-semiemp}, is used to compute the C$_4$ coefficient of Eq. \ref{eq:capture}.
    Finally, the red curve shows the final values obtained by applying the steric factor, described in the text. 
    The magenta triangle shows the measurements of the rate constant obtained at 340 K by \citet{meot1986ion}.
    \textit{Right panel:}  Rate constants of reaction [9b], CH$_3$OH$_{2}^{+}$ + HNC, as a function of the temperature for the products CH$_3$CNH$^+$ + H$_{2}$O (green) and CH$_3$OH + HCNH$^+$ (blue), respectively, plus the back dissociation rate (red).}
    \label{fig:rate-reac9}
\end{figure*}
The derived steric factor, which are given by the ratio between the number of attractive configurations and the total number of configurations, is equal to 0.58 for the formation of INT1, INT2 and INT3.
By multiplying the previous rate constant by this factor we obtained a rate of $k=6.5\times10^{-11}$ cm$^3$ s$^{-1}$ at 340 K, reaching a factor 2 of difference with the experimental value.

The final rate constant as a function of the temperature is shown in Fig.\ref{fig:rate-reac9}. 
We note that the back dissociation plays an important role in determining the trend of the rate constant: the larger the temperature, the lower the rate constant due to the increasing role of the back dissociation. 

\vspace{0.3cm}
\noindent
\textit{Reaction [9b]: CH$_3$OH$_{2}^{+}$ + HNC $\rightarrow$ CH$_{3}$CNH$^{+}$ + H${_2}$O}

As discussed before, the reaction between protonated methanol and HNC leads to the formation of protonated methyl cyanide.
Using the same methodology described for the reaction [9a], we computed the C$_4$ coefficient and the steric factor which is found to be equal to 0.69.
The main reaction channel is the one leading to CH$_3$OH + HCNH$^+$ with a rate constant of $1.04\times10^{-9}$ cm$^3$ s$^{-1}$. 
The channel of formation of CH$_3$CNH$^+$ + H${_2}$O has instead a minor contribution (BR equal to 0.01). 

Figure \ref{fig:rate-reac9} shows the rate constant as a function of the temperature of the two reactive channels. 
The contribution of the back dissociation is negligible in this case.

\subsection{Reactions [10]: \texorpdfstring{CH$_3$CNH$^+$ + NH$_3$ or e$^-$ $\rightarrow$ CH$_{3}$CN + NH$_{4}^{+}$ or H}{reaction10} \label{subsec:results-reaction10}}

\begin{figure*}
    \includegraphics[width=8.2cm]{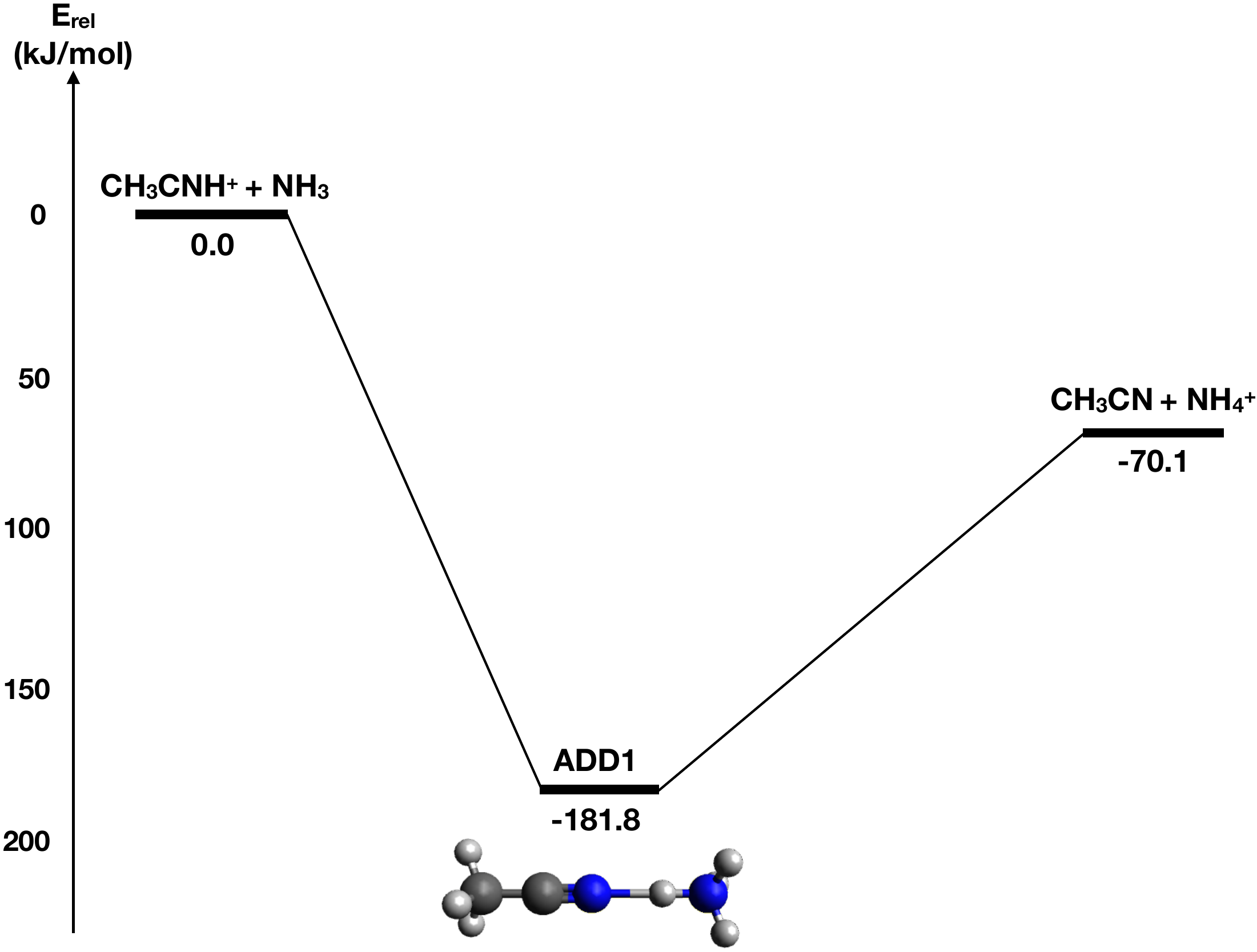}
    \includegraphics[width=8.2cm]{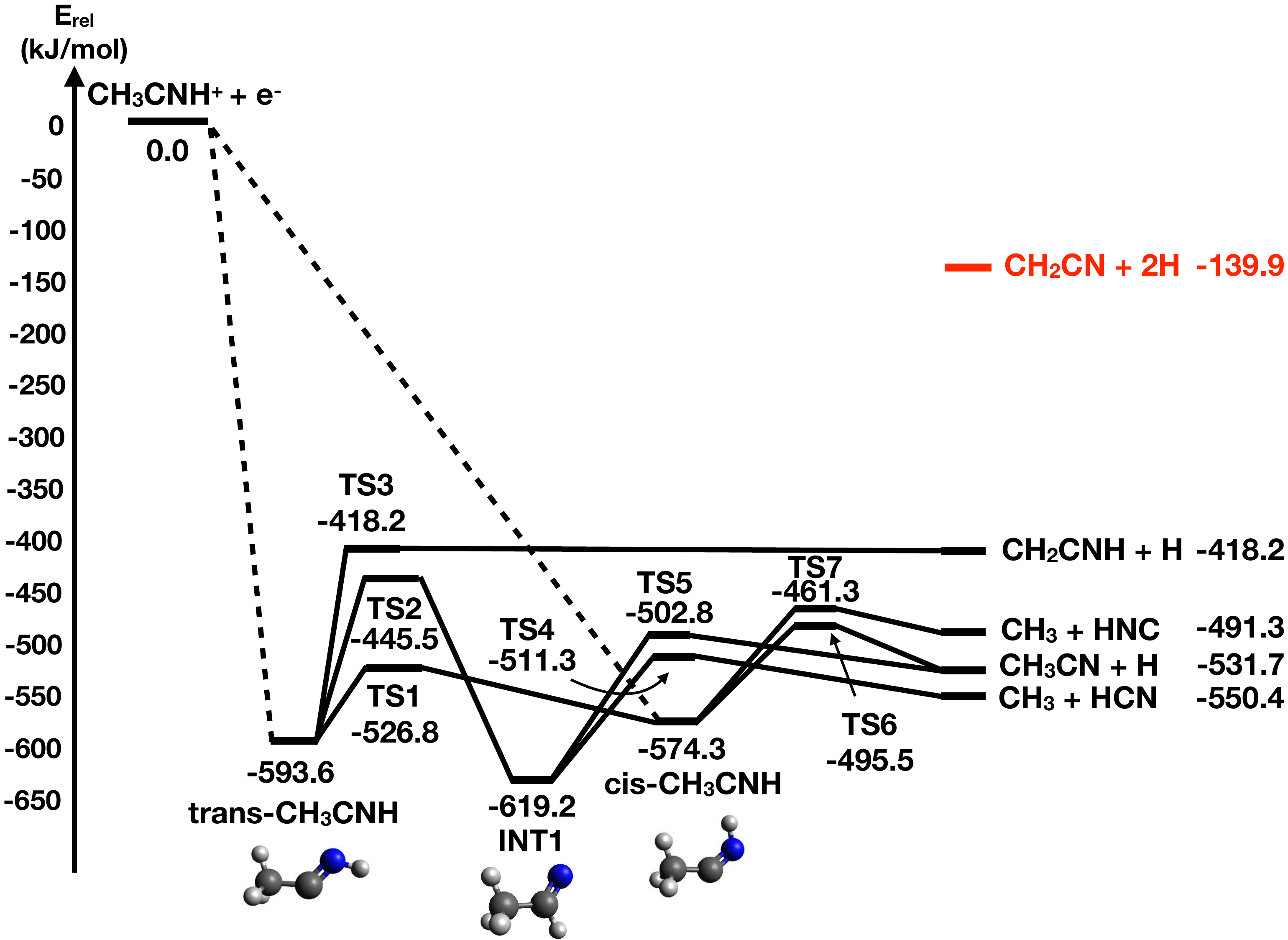}
    \caption{PES of the reaction [10a] (\textit{left panel}) and [10b] (\textit{right panel}) of Table \ref{tab:overview-reactionlist}, CH$_3$CNH$+$ + NH$_3$ or e$^-$.
    The plots show the results obtained at the CCSD(T)/aug-cc-pVTZ//B3LYP/aug-cc-pVTZ level of theory. }
    \label{fig:pes-reac10}
\end{figure*}

As for the reactions [9a] and [9b], QM computations were not previously available in the literature.
Since both reactions are important in the formation of methyl cyanide, we discuss their case in detail here.
We anticipate the results for the reader not interested in such details.

The protonated methyl cyanide exchanges the proton with ammonia forming methyl cyanide with a rate constants equal to $8.7 \times 10^{-10}$ cm$^3$ s$^{-1}$, constant in the 10--300 K interval.
On the contrary, only about 7\% of the protonated methyl cyanide electron recombination produces methyl cyanide.
The dependence of the rate constants goes with the -1.03 dependence on the temperature.

\subsubsection{PES of reactions [10]}

\noindent
\textit{Reactions [10a]: CH$_3$CNH$^+$ + NH$_3$ $\rightarrow$ CH$_{3}$CN + NH$_{4}^{+}$} \label{subsubsec:PES-reaction10a}

The schematic PES of reaction [10a] is reported in Fig. \ref{fig:pes-reac10}, left panel.
The interaction between protonated methyl cyanide and ammonia gives rise, in a barrier-less process, to an adduct (ADD1) more stable than the reactants by 181.8 kJ mol$^{-1}$. 
After that, the dissociation of the adduct into methyl cyanide and ammonium ion is exothermic by 70.1 kJ mol$^{-1}$, in perfect agreement with the differences of proton affinities of ammonia and CH$_3$CN retrieved from the NIST database \citep{NIST_ASD}. 

\vspace{0.3cm}
\noindent
\textit{Reaction [10b]: CH$_3$CNH$^+$ + e$^-$ $\rightarrow$ CH$_{3}$CN + H} \label{subsubsec:PES-reaction10b}

Figure \ref{fig:pes-reac10}, right panel, shows the PES of reaction [10b].
Please note that only the most relevant channels are reported in the figure. 
Following the recombination of CH$_3$CNH$^+$ with an electron two intermediates can be formed, trans-CH$_3$CNH or cis-CH$_3$CNH with an exothermicity of -593.6 kJ mol$^{-1}$ and -574.3 kJ mol$^{-1}$, respectively. 
These two intermediates can isomerize between each other through TS1, overcoming a barrier of 66.8 kJ mol$^{-1}$ and 47.5 kJ mol$^{-1}$, respectively. 
The trans-CH$_3$CNH isomer can form the CH$_{2}$CNH + H products, localised at -418.2 kJ mol$^{-1}$ with respect to the reactants, via TS3 (barrier of 174.4 kJ mol$^{-1}$); otherwise it can isomerize to INT1 (-619.2 kJ mol$^{-1}$) by transferring an hydrogen from the N atom to the C atom via TS2 (barrier of 148.1 kJ mol$^{-1}$). 
The fission of a C-H bond in INT1 leads to the formation of CH$_3$CN + H, localised at -531.7 kJ mol$^{-1}$ below the reactant energy asymptote, trough TS5 (barrier of 116.4 kJ mol$^{-1}$), or the fission of the C-C bond leads to the formation of CH$_3$ + HCN localised at -531.7 kJ mol$^{-1}$ with respect to the reactants, through TS4 (barrier of 107.9 kJ mol$^{-1}$). 
An H-displacement in cis-CH$_3$CNH can also lead to the formation of CH$_3$CN overcoming a barrier of 78.8 kJ mol$^{-1}$ (TS6), otherwise it can form CH$_3$ + HNC through TS7 (barrier of 113.0 kJ mol$^{-1}$) after the fission of the C-C bond. 
Further fragmentation of CH$_3$CN or CH$_{2}$CNH are possible considering the amount of energy left after their formation, leading to CH$_{2}$CN + H + H which is still exothermic by 139.9 kJ mol$^{-1}$. 
The connections are not reported in the PES but this possibility is taken into account in the kinetics computations. 

\subsubsection{Kinetics of reaction [10]}

\noindent
\textit{Reactions [10a]: CH$_3$CNH$^+$ + NH$_3$ $\rightarrow$ CH$_{3}$CN + NH$_{4}^{+}$}

Since the back dissociation of reaction [10a] is negligible, the rate constant is essentially independent from temperature and equal to the capture rate constant $8.7 \times 10^{-10}$ cm$^3$ s$^{-1}$ in the studied temperature range.
This value is in agreement with the lower limit provided by \citet{hemsworth1974rate}, $\sim 1\times 10^{-9}$ cm$^3$ s$^{-1}$.

\vspace{0.3cm}
\noindent
\textit{Reaction [10b]: CH$_3$CNH$^+$ + e$^-$ $\rightarrow$ CH$_{3}$CN + H}

In this reaction the two cis-CH$_3$CNH and trans-CH$_3$CNH intermediates can be initially formed. 
Starting from each of them we solved the master equation in order to derive the BF of formation of the products and we weighted the results for the density of states of the first two intermediates. 
Due to the impossibility to derive a capture rate constant, we did not compute the rate constants of formation of each product. 
From our calculations we obtained that 60\% of the products preserve the carbon chain, of which 53\% is CH$_{2}$CNH and 7\% is CH$_3$CN, while 40\% of the products break the C-C or C-N bond, forming CH$_3$ + HNC (39\%) or CH$_3$ + HCN (1\%). 
These results are in good agreement with the experimental study of \citet{geppert2007formation}, which found that 65\% of the products preserve the carbon-chain and 35\% fragment in smaller species.
We then weighted the rate constant measured by \citet{mclain2009flowing} with our calculated BFs to derive the rate constant of formation of each product. We kept the temperature dependence of -1.03 measured by \citep{mclain2009flowing} for non-deuterated CH$_3$CNH$^+$. 
The obtained values are reported in Table \ref{tab:revised-network} for the different major products.
Note that, since CH$_{2}$CNH is likely further fragmented \citep[as also suggested by][]{loison2014interstellar}, we added the rate constants of the channel that produces CH$_{2}$CNH + H to that of the channel CH$_{2}$CN + H + H.


%
\renewcommand{\arraystretch}{1.8}
\begin{landscape}
\begin{table}
    \begin{center}
    \caption{Revised network for the formation of methyl cyanide in the gas-phase.
    For each reaction (column 1), the rate constants are given in terms of $\alpha$ (in cm$^3$ s$^{-1}$); column 2), $\beta$ (column 3) and $\gamma$ (column 4).
    They follow the formalism of Eq. \ref{eq:Arrhenius}, except for reaction [3] where the Ion-pol formalism is employed instead \citep{wakelam2012kinetic}.
    The rate constants are valid for the range of temperature reported in column 5.
    Note that the number of each reaction refers to those in Table \ref{tab:overview-reactionlist}.
    The reactions preceded by a star are those revised based on either a careful literature search or our new QM computations, presented in Sec. \ref{sec:results}; those preceded by a plus are taken from the databases, but they would need to be verified; the remaining reactions are confirmed after our critical review of Sec. \ref{sec:overview-gas-routes}.
    For comparison, columns 6 to 11 report the values listed in the KIDA \citep{wakelam2012kinetic} and UDfA \citep{mcelroy2013umist} databases, respectively.}
	\label{tab:revised-network}
	\begin{tabular}{ccrcl|cccc||ccc|ccc}
		\hline
 \multicolumn{9}{c||}{Revised network} & \multicolumn{3}{c|}{KIDA} & \multicolumn{3}{c}{UDfA} \\
\multicolumn{5}{c}{Reaction} & $\alpha$ & $\beta$ & $\gamma$ & T [K]  & $\alpha$ & $\beta$ & $\gamma$ & $\alpha$ & $\beta$ & $\gamma$ \\
	\hline
		\hline
  \multicolumn{15}{l}{\textit{(i) Routes to CH$_3$CN$^+$}}\\
3 &+& NH$_{2}$CHO + C$^+$ & $\rightarrow$ & CH$_{3}$CN$^{+}$ + O &  $1.67\times10^{-1}$ & $1.55\times10^{-9}$ & 6.62 & 10-300   & $1.67\times10^{-1}$ & $1.55\times10^{-9}$ & 6.62  & - & - & - \\
\hline
\multicolumn{15}{l}{\textit{(ii) Routes to CH$_3$CNH$^+$}}\\
5a& & CH$_{3}$CN$^{+}$ + H$_{2}$ & $\rightarrow$ & CH$_{3}$CNH$^+$ + H & $5.7\times10^{-10}$ & 0.0 & 0.0 & 10-300  & - & - & - & $5.7\times10^{-10}$ & 0.0 & 0.0\\
5b& & CH$_{3}$CN$^{+}$ + CH$_4$ & $\rightarrow$ & CH$_{3}$CNH$^+$ + CH$_3$ & $1.7\times10^{-9}$ & 0.0 & 0.0& 10-300  & - & - & - & $1.7\times10^{-9}$ & 0.0 & 0.0 \\
6 & & C${_2}$H${_7}{^+}$ + HCN & $\rightarrow$ & CH$_{3}$CNH$^{+}$ + CH${_4}$ & $2.2\times10^{-10}$ & -0.5 & 0.0 & 10-300   & - & - & - & $2.2\times10^{-10}$ & -0.5 & 0.0 \\
7a&*& CH${_3}{^+}$ + HCN & $\rightarrow$ & CH$_{3}$CNH$^{+}$ + h$\nu$ & $5.7\times10^{-9}$ & -0.5 & 0.0 & 10-300  & $9.0\times10^{-9}$ & -0.5 & 0.0 & $9.0\times10^{-9}$ & -0.5 & 0.0 \\
8 &*& CH${_4}{^+}$ + HCN & $\rightarrow$ & CH$_{3}$ + HCNH$^{+}$ & $3.9\times10^{-9}$ & 0.0 & 0.0 &  10-300  & $3.3\times10^{-9}$ & 0.0 & 0.0 & - & - & - \\
  &*&                    &               & CH$_3$CNH$^+$ + H     & $3.2\times10^{-11}$& 0.0 & 0.0 &  10-300  & $3.3\times10^{-9}$ & 0.0 & 0.0 & - & - & - \\
9b&*& CH$_{3}$OH${_2}{^+}$ + HNC & $\rightarrow$ & CH$_{3}$CNH$^{+}$ + H${_2}$O & $3.0\times10^{-11}$  & 0.0 & 0.0 & 10-300 & - & - & - & - & - & - \\
  &*& & $\rightarrow$ & CH$_{3}$OH + HCNH$^{+}$ & $1.0\times10^{-9}$  & 0.0 & 0.0 &  10-300 & - & - & - & - & - & - \\
  \hline
\multicolumn{15}{l}{\textit{(iii) Routes to CH$_3$CN}}\\
10a&*&CH$_{3}$CNH$^{+}$ + NH$_3$ & $\rightarrow$ & CH$_{3}$CN + NH$_{4}^{+}$ & $8.7\times10^{-10}$ & 0.0 & 0.0 & 10-300 & - & - & - & - & - & - \\
10b&*& CH$_{3}$CNH$^{+}$ + e$^-$ & $\rightarrow$ & CH$_{3}$CN + H  & $2.2\times10^{-8}$ & -1.03 & 0.0 & 10-300 & $1.3\times10^{-7}$ & -0.5 & 0.0 & $5.3\times10^{-7}$ & -0.7 & 0.0 \\
&*& & $\rightarrow$ & CH${_2}$CN + H + H & $1.8\times10^{-7}$  & -1.03 & 0.0 & 10-300 & $8.8\times10^{-8}$ & -0.5 & 0.0 & - & - & -\\
&*& & $\rightarrow$ & CH$_{3}$ + HCN  & $3.4\times10^{-10}$ & -1.03 & 0.0 & 10-300 & $6.0\times10^{-8}$ & -0.5 & 0.0 & - & - & -\\
&*& & $\rightarrow$ & CH$_{3}$ + HNC  & $1.4\times10^{-7}$ & -1.03 & 0.0 & 10-300 & $6.0\times10^{-8}$ & -0.5 & 0.0 & $2.8\times10^{-7}$ & -0.7 & 0.0 \\
11b&+& CH$_{3}$ + CN$^-$ & $\rightarrow$ & CH$_{3}$CN + e$^-$ & $1.0\times10^{-9}$ & 0.0 & 0.0 & 10-300  & $1.0\times10^{-9}$ & 0.0 & 0.0 & $1\times10^{-9}$ & 0.0 & 0.0 \\
    \hline
	\end{tabular}
	\end{center}
\end{table}
\end{landscape}


\section{The revised gas-phase reaction network for the methyl cyanide formation} \label{sec:revised-network}

Figure \ref{tab:revised-network} shows the new scheme of the gas-phase reaction network leading to the formation of methyl cyanide.
Table \ref{tab:revised-network} summarizes the reactions, products and rate constants of the new revised network for the formation of methyl cyanide.
The rate constants are given in terms of the parameters $\alpha$, $\beta$ and $\gamma$, following the Eq. \ref{eq:Arrhenius}, which is commonly used in the astrochemical databases and models. 
Please note that $\alpha$, $\beta$ and $\gamma$ are the results of the fits of the rate constants as a function of the temperature, in the temperature range marked in the table. 
Therefore, the $\gamma$ parameter is sometimes different from zero even in absence of an activation barrier, because it just makes the analytical fit better, as it often occurs in literature.
It is important to emphasize that, in those cases, $\gamma$ must not be considered as an activation barrier.

The new network consists of ten reactions, forming methyl cyanide or competitive products.
The changes with respect to the original one, summarised in Tab.\ref{tab:overview-reactionlist}, are:
\begin{itemize}
    \item [-] Reactions [1], [2] and [4] have been removed because they do not form CH$_3$CN$^+$, as incorrectly reported in KIDA and UDfA, but its isomer, CH$_{2}$CNH$^+$.
    \item [-] Reaction [3] has not been revised, since it is likely negligible in the formation of methyl cyanide.
    \item [-] Reactions [5a] and [5b] are verified based on the literature review. Therefore, they have been included.
    Note that reaction [5a] is reported in UDfA but not in KIDA, while reaction [5b] is not reported in any of the two databases. 
    \item [-] Reactions [6] is also verified based on the literature review. It has been included in the network since it was correctly reported in the UDfA but it was not present in KIDA. 
    \item [-] Reaction [7a] has been revised and a rate constant smaller by a factor 2 than the one reported in KIDA and UDfA has been adopted.
    \item [-] Reaction [7b] has been removed because it forms CH$_{3}^{+}$ + HCN and not CH$_3$CNH$^+$ as incorrectly reported in KIDA and UDfA.
    \item [-] Reaction [8] mainly forms CH$_3$ + HCNH$^+$ rather than CH$_3$CNH$^+$, and the rate constant for the formation of CH$_3$CNH$^+$ is 100 times lower than that reported in KIDA.
    \item [-] Reaction [9a] has been removed because it does not form CH$_3$CNH$^+$, as incorrectly reported in UDfA, but its isomer, CH$_3$NCH$^+$.
    \item [-] Reaction [9b] has been added (it is not present in KIDA or UDfA) as it forms CH$_3$CNH$^+$, according to our new QM computations.
    \item [-] Reaction [10a] has been added (it is not present in KIDA or UDfA) as it forms CH$_3$CN, according to our new QM computations. 
    \item [-] Reaction [10b] has been modified following our new QM computations of the BFs. Specifically, the BF of formation of CH$_3$CN has been decreased from 39\% to 6.5\%. 
    \item [-] Reaction [11a] has been removed since it forms CH$_2$ + HCN and not CH$_3$CN as incorrectly reported in KIDA and UDfA.
    \item [-] Reaction [11b] has not been revised, since it is likely negligible in the formation of methyl cyanide.
    \item [-] Reaction [12] has been removed since it does not form CH$_3$CN, contrary to what reported in KIDA and UDfA. 
\end{itemize}

To summarise, the formation of methyl cyanide mainly follows three steps (see also Tab. \ref{tab:revised-network}): 
\begin{itemize}
    \item [(1)] formation of CH$_3$CN$^+$ from the reaction of NH$_2$CHO with C$^+$. However, this step is found to be negligible for the formation of CH$_3$CN following the model described in Sec. \ref{sec:astro-modeling}. 
    \item [(2)] formation of CH$_3$CNH$^+$ from the reaction of the CH$_3^+$ , CH$_4^+$, CH$_3$OH$_2^+$ and  C$_2$H$_7^+$ ions with HCN (or HNC), or from the hydrogen abstraction of CH$_3$CN$^+$ from H$_2$ or CH$_4$;
    \item [(3)] formation of CH$_3$CN either from CH$_3$CNH$^+$, via the recombination with e$^-$ or the proton transfer to NH$_3$, or via the reaction of CH$_3$ with CN$^-$.
\end{itemize}



\begin{figure}
    \includegraphics[width=8.5cm]{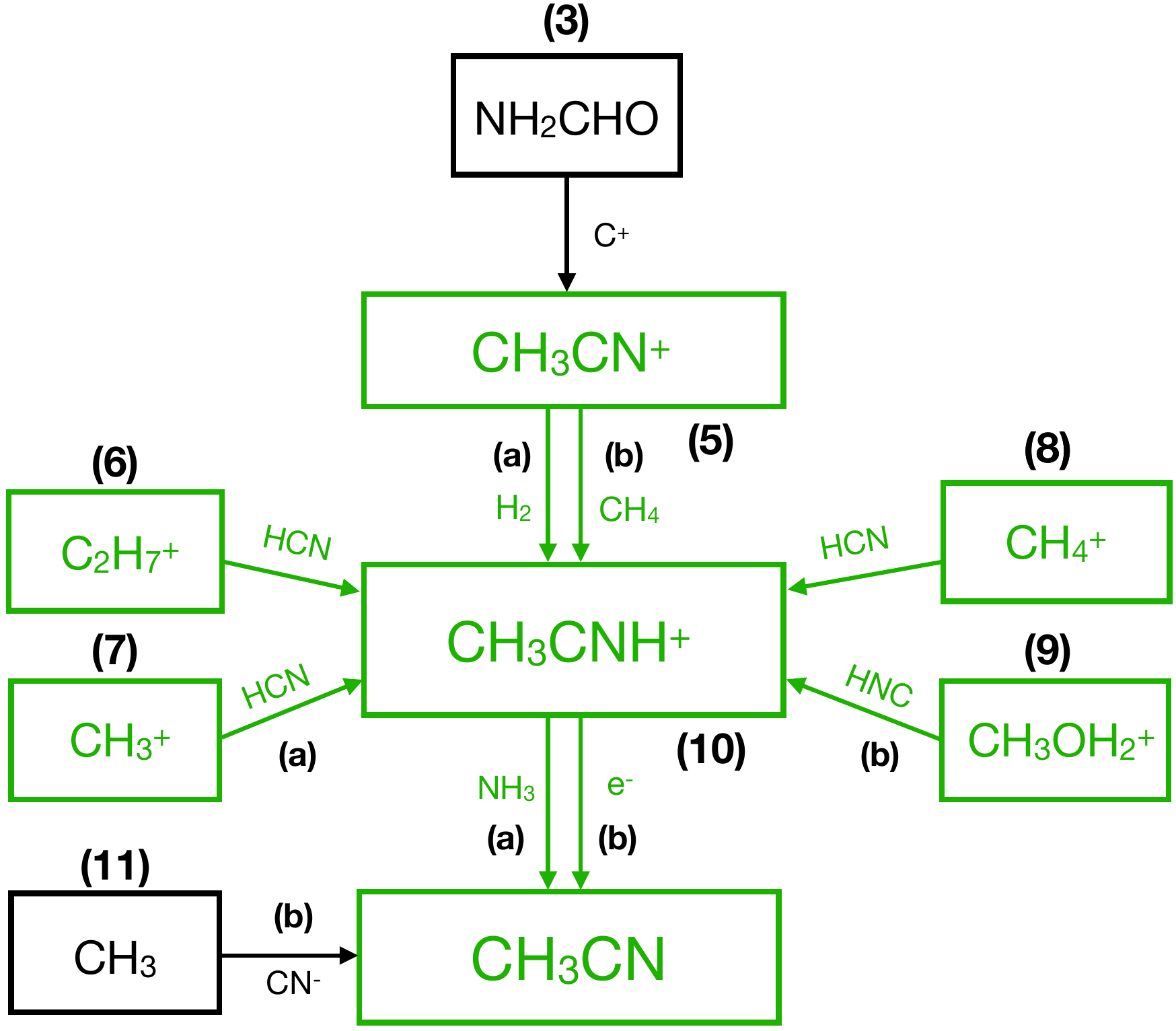}
    \caption{Scheme of the new gas-phase reaction network leading to the formation of methyl cyanide, following the revision summarised in Sec. \ref{sec:revised-network}.
    The numbers on top of each box are those in Tab. \ref{tab:revised-network}.
    The green boxes mark the reactions which have been verified to be correct while the black boxes mark reactions that would need further investigations (but are unlikely important in the formation of methyl cyanide).}
    \label{fig:overview-reactionscheme-AR}
\end{figure}

\section{Astrochemical modeling} \label{sec:astro-modeling}

In order to understand which gas-phase reactions of the network described in Sec. \ref{sec:overview-gas-routes} (see also Tab. \ref{tab:overview-reactionlist} and Fig. \ref{fig:overview-reactionscheme}) have the largest impact in the formation of CH$_3$CN, we ran simulation with the astrochemical model described in Sec. \ref{subsec:astro-model-description}.
We considered two general cases with different conditions and where methyl cyanide has been observed, as discussed in Sec. \ref{sec:discussion}.
Specifically, we simulated the chemical evolution of a cold molecular cloud, described in Sec. \ref{subsec:astro-model-cloud}, and a warm protostellar molecular shock, described in Sec. \ref{subsec:astro-model-shock}, respectively.
We emphasize that the goal of the modeling and the focus of the results discussion are to understand the importance of the various reactions in forming methyl cyanide in different conditions and whether the new reaction network is capable to approximately reproduce the observed abundance of methyl cyanide.

\subsection{Description of the astrochemical model} \label{subsec:astro-model-description}
We used the code MyNahoon, which is a time-dependent gas-phase chemical model based on the Nahoon code originally developed by \cite{wakelam2005estimation,wakelam2010sensitivity}.
Briefly, the code computes the abundances of gaseous species as a function of time for a given  temperature $T$ and density $n_{H}$, and other standard parameters detailed below.
The code does not consider reactions occurring on the grain surfaces, except for the formation of H$_{2}$ molecules. 
The simulation starts from an initial composition that depends on the object to simulate and that also defines the overall gaseous elemental abundances.
It then follows the chemical evolution with time until the age, which also depends on the simulated object, is reached.

For the gas-phase reaction network, we used the GRETOBAPE network, described in \cite{tinacci2023-gretobape}, and the new network of Tab. \ref{tab:revised-network} for the methyl cyanide formation (see also Fig. \ref{fig:overview-reactionscheme-AR}).
Briefly, the GRETOBAPE network is based on the KIDA2014 network \cite{wakelam20152014}, upgraded following studies from our and other groups \cite[e.g.]{loison2014interstellar,vazart2016state,balucani2015formation,skouteris2018genealogical,vazart2020gas,neufeld2015sulphur,codella2020seeds,blazquez2020gas}.
Very important, the GRETOBAPE network is cleaned by endothermic reactions that were erroneously present in the original KIDA2014 network.

\subsection{Cold molecular cloud} \label{subsec:astro-model-cloud}
\subsubsection{Modeling and adopted parameters}

To simulate a generic molecular cloud, where methyl cyanide has been detected, we assumed the following parameters: 
gas temperature $T=10$ K,  
H-nuclei density $N_H=2\times 10^4$ cm$^{-3}$, 
cosmic-ray (CR) ionization rate $\zeta_{CR}=3\times 10^{-17}$ s$^{-1}$, 
visual extinction A$_v$=20 mag, 
gas-to-dust ratio in mass equal to 0.01, 
dust grain radius $a_d$=0.1 $\mu$m and  grain density 3 g cm$^{-3}$.  

We aimed at simulating the evolution of a typical translucent cloud into a molecular cloud.
For this reason, as initial conditions, all elements with an ionisation potential below 13.6 eV are ionised (C, S, Si, P...) while the elements with larger ionisation potential are neutral (N and O). 
Hydrogen is assumed to be fully molecular.
The adopted elemental abundances are summarized in Table \ref{tab:astro-cold-initial-abd}. 

\begin{table}
    \begin{center}
	\caption{Initial elemental abundances relative to H nuclei adopted for the cold molecular cloud modeling. The abundances are adapted from \citealt{jenkins2009unified}.}
	\label{tab:astro-cold-initial-abd}
	\begin{tabular}{cc|cc}
    \hline
    \hline
    Element & Abundance & Element & Abundance \\
    \hline
He      & $9.0 \times 10^{-2}$ & P$^+$   & $2.0 \times 10^{-10}$ \\
C$^+$   & $1.7 \times 10^{-5}$ & Na$^+$  & $2.0 \times 10^{-9}$ \\
O       & $2.6 \times 10^{-5}$ & Mg$^+$  & $7.0 \times 10^{-9}$ \\
N       & $6.2 \times 10^{-6}$ & Fe$^+$  & $3.0 \times 10^{-9}$ \\
S$^+$   & $8.0 \times 10^{-8}$ & Cl$^+$  & $1.0 \times 10^{-9}$ \\
Si$^+$  & $8.0 \times 10^{-9}$ & F$^+$   & $1.0 \times 10^{-9}$ \\
	\hline
	\end{tabular}
	\end{center}
\end{table}
%

\subsubsection{Results of the modeling} \label{subsubsec:modelresults-mc}

\noindent

\begin{figure}
	\includegraphics[width=8.3cm]{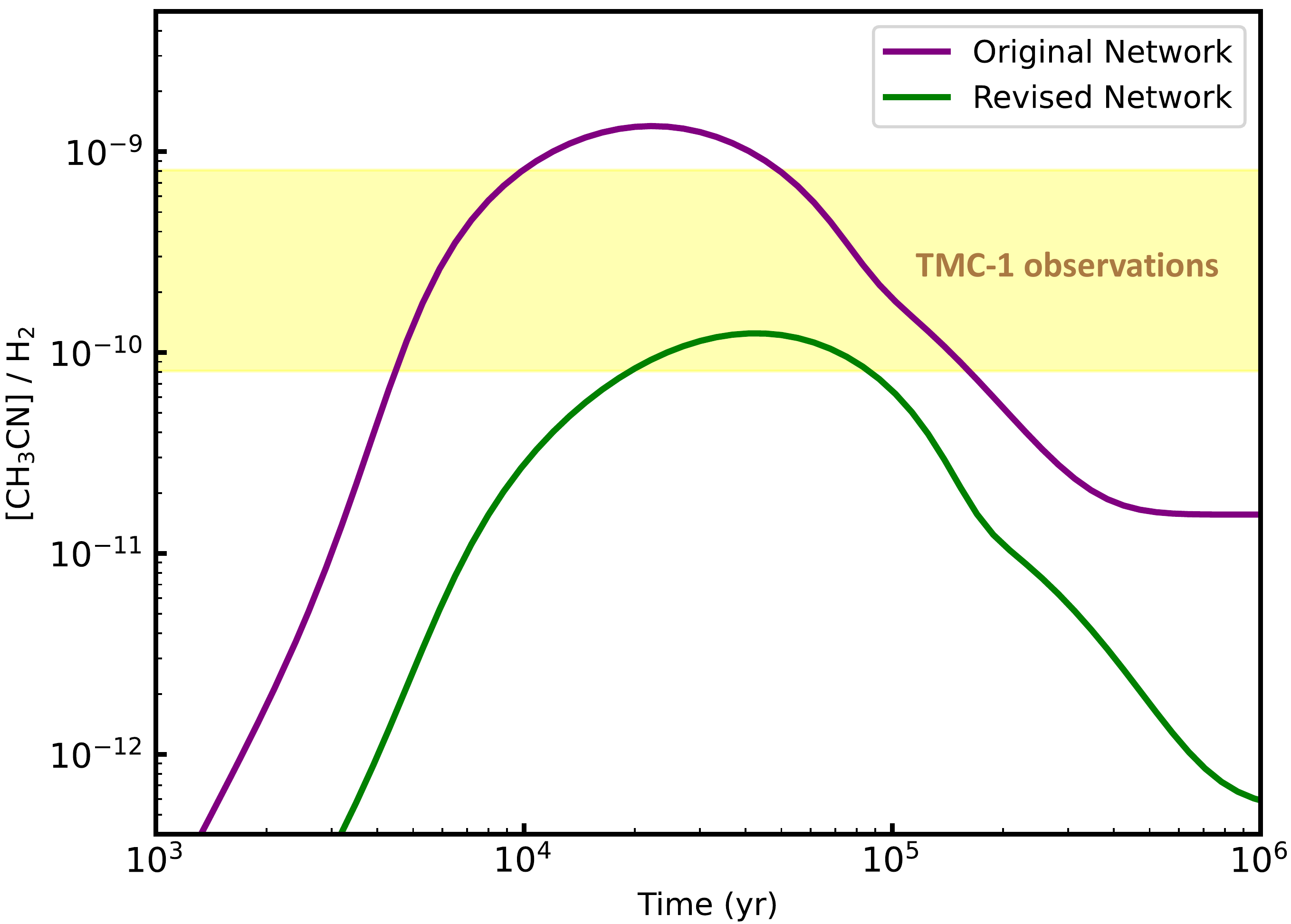}
    \caption{Abundance (with respect to H$_2$) of methyl cyanide as a function of time, using the original network (purple) and the revised network (green). 
    The yellow band shows the measured abundance of methyl cyanide in the TMC-1 cold cloud (see Table \ref{tab:detections}).}
    \label{fig:model-ch3cn-abundance-cloud}
\end{figure}

In order to see the effect of the new reaction network, we ran the cold cloud model using both the old and the new networks. 
Figure \ref{fig:model-ch3cn-abundance-cloud} shows the methyl cyanide abundance as a function of time, using the revised and the original reaction networks, respectively. 
As predicted by other astrochemical models for several C-bearing species, the CH$_3$CN abundance has a peak at $\sim 10^5$ yr, when ionised carbon becomes neutral but is not yet mostly locked into CO. 

With both the networks, the methyl cyanide formation follows the two steps summarised in Sec. \ref{sec:revised-network}: 
(1) the CH$_3$CNH$^+$ formation, dominated by the radiative association of CH$_{3}^{+}$ with HCN, followed by 
(2) the formation of CH$_3$CN, dominated by the electron recombination of CH$_3$CNH$^+$. 
In other words, the major reactions involved in the methyl cyanide formation in cold molecular clouds are [7a] and [10b]. 

That said, it is important to emphasize that, in the revised network, the rate constants of these two reactions have been modified.
The rate constant of reaction [7a] has been decreased from $9.0 \times 10^{-9}$ cm$^3$ s$^{-1}$ to $5.7 \times 10^{-9}$ cm$^3$ s$^{-1}$ (see Sec. \ref{subsec:overview-reaction7}), and the BF of formation of CH$_3$CN in reaction [10b] has been lowered from 39\% to 7\% (see Sec. \ref{subsec:results-reaction10}). 
The new values adopted for reactions [7a] and [10b] affect the CH$_3$CN abundance, which decreases by one order of magnitude with respect to the one obtained using the original network. 

Finally, we searched for the impact of the new reaction network on species other than CH$_3$CN.
The most affected ones are CH$_3^+$, CH$_3$, HCN and HNC, which in turn are involved in the formation of C$_2$H$_2^+$, C$_2$H$^+$, C$_3$N and CH$_4^+$. However, the abundances predicted using the new network differ by less than one order of magnitude with respect to those predicted by the original network.

\subsection{Warm molecular protostellar shock} \label{subsec:astro-model-shock}

\subsubsection{Modeling and adopted parameters}\label{subsubsec:astro-shock}

Young protostars are known to have spectacular ejections of matter which cause shocks when they hit the surrounding quiet environment.
From the chemical point of view, the passage of the shock has two major effects: (i) it heats and compresses the gas and (ii) species frozen on the icy grain mantles are sputtered and injected into the gas phase.
As a result, molecular shocks are warm ($\sim$80--100 K) and dense ($\geq 10^5$ cm$^{-3}$) regions with much larger abundances of some species with respect to cold molecular clouds.

In order to simulate the chemical composition of the gas after the passage of a shock, we adopt a two-step procedure where we first compute the chemical abundances of the cold molecular cloud (previous subsection) and then we suddenly (i) increase the gas temperature (but not the dust one, which remains roughly the same before the shock event) and the H nuclei number density and (ii) (artificially) inject into the gas large abundances of species that are known to be present on the grain icy mantles, because of the sputtering and shattering of the grains.
Although the physics is slightly different, a similar modeling can be also considered a fair enough simulation of what happens in hot cores and hot corinos, which are regions heated by the central forming star and where the grain icy mantles sublimate \citep[e.g. see the recent review by][]{Ceccarelli2023-PP7}.

In practice, our model has two steps, as follows.\\
\underline{Step 1:} We first compute the steady-state chemical composition of a cold
molecular cloud at 10 K with a H nuclei number density of $2\times 10^4$ cm$^{-3}$ and with the parameters described in Sec. \ref{subsec:astro-model-cloud}.\\
\underline{Step 2:} We increase the gas temperature to 90 K, H nuclei number density to
$8 \times 10^5$ cm$^{-3}$ and we inject into the gas the grain-mantle species reported in Table \ref{tab:model-parameters} with the abundances listed in the table.
We then leave the chemical composition evolve for $10^4$ yr.

The values adopted for the physical parameters and abundances of the injected species are based on the astronomical observations of the prototypical molecular outflow shock L1155-B1, for which our group carried the same model described here in previous works \cite[e.g.][]{codella2017seeds,podio2017silicon,codella2020seeds}.
Please note, though, that the modeling also approximately describes the chemical evolution of hot corinos, where similarly to the molecular shock, the components of the grain mantles are injected into the gas-phase (because of their thermal sublimation).

\begin{table}
    \centering
    \begin{tabular}{c|c}
        \hline
        \hline
        \multicolumn{2}{c}{Physical parameters of the shocked gas} \\
        Parameter & L1157-B1 value \\
        \hline
        n$_{H_2}$    [cm$^{-3}$] & $4\times 10^5$      \\
        T            [K]         & 90                  \\
        $t_{shock}$ [yr]         &  200-3000               \\
        $\zeta_{CR}$ [s$^{-1}$]  & $6\times 10^{-16}$ \\
        \hline
        \hline
        \multicolumn{2}{c}{Abundances (wrt H) of the injected species} \\
        Species & L1157-B1 value \\
        \hline
        H$_{2}$O    & $1\times 10^{-4}$ \\
        CO$_{2}$    & $3\times 10^{-5}$ \\
        CO    & $1\times 10^{-4}$ \\
        CH$_3$OH  & $8\times 10^{-6}$ \\
        NH$_3$    & $5.6\times 10^{-5}$ \\
        H$_{2}$CO   & $1\times 10^{-6}$ \\
        OCS       & $2\times 10^{-6}$ \\
        SiO       & $1\times 10^{-6}$ \\
        Si        & $1\times 10^{-6}$\\
        CH$_3$CH$_{2}$ & $4\times 10^{-8}$  \\
        CH$_3$CH$_{2}$OH & $6\times 10^{-8}$ \\
        SiH$_4$  & $1\times 10^{-7}$ \\
        \hline
    \end{tabular}
    \caption{Parameters used to model the passage of the shock (Sect. \ref{subsec:astro-model-shock}).
    The upper half table lists the adopted H$_{2}$ density, n$_{H_2}$, temperature, $T$, age, $t_{shock}$ and cosmic-ray ionisation rate, $\zeta_{CR}$, of the gas after the passage of the shock.
    The lower half table lists the adopted abundance of the species injected into the gas-phase from the grain mantle after the passage of the shock.
    }
    \label{tab:model-parameters}
\end{table}

\subsubsection{Results of the modeling} \label{subsubsec:astro-shock-results}

\begin{figure}
	\includegraphics[width=8.5cm]{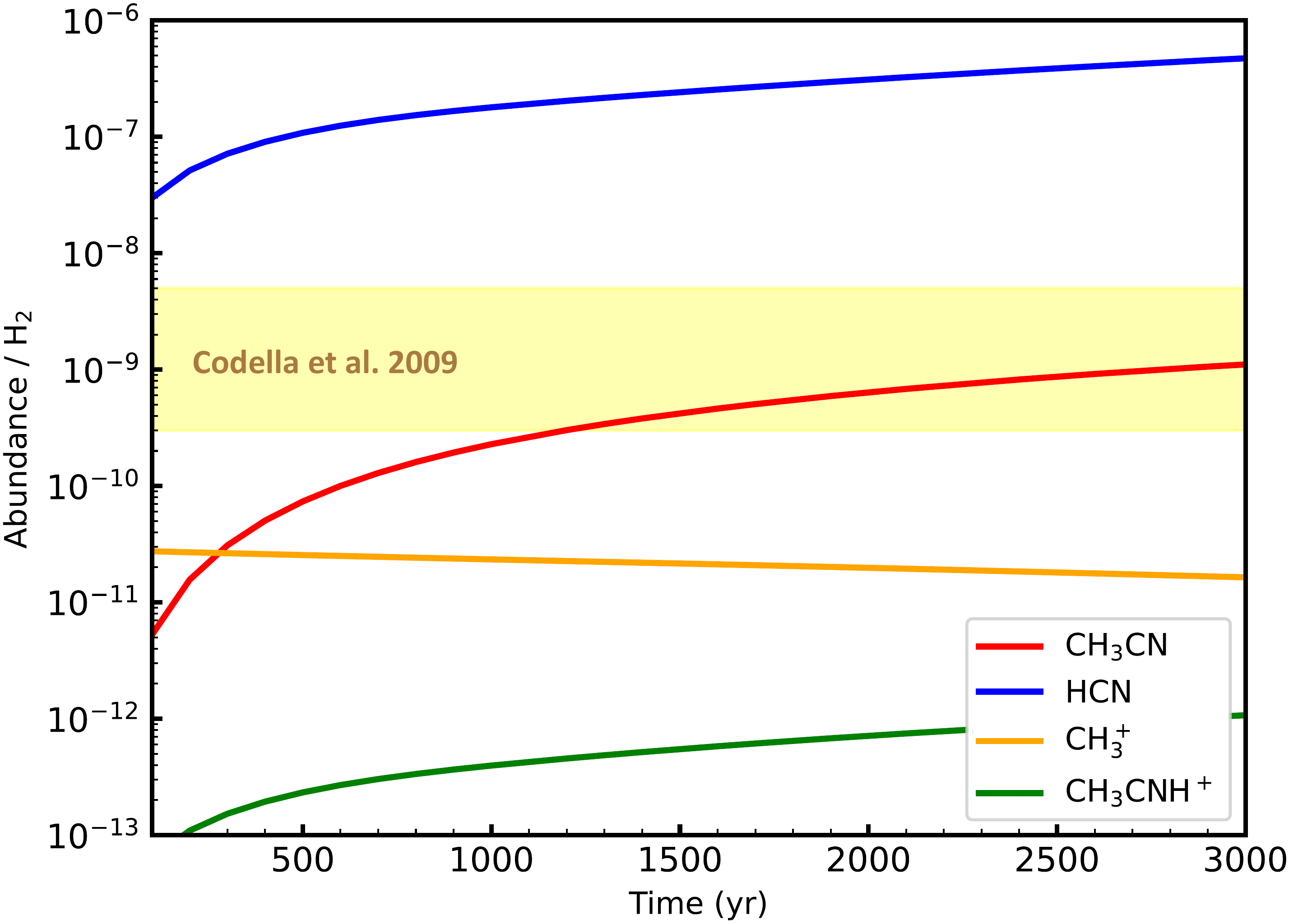}
    \caption{Abundances (with respect to H$_2$) of methyl cyanide and other species involved in its formation as a function of time from the passage of the shock: CH$_3$CN (red), HCN (blue), CH$_3^+$ (orange) and CH$_3$CNH$^+$ (green).
    The yellow band shows the CH$_3$CN abundance observed towards the L1157-B1 shocked region (see Table \ref{tab:detections}).}
    \label{fig:model-ch3cn-abundance-shock}
\end{figure}

Figure \ref{fig:model-ch3cn-abundance-shock} shows the post-shock evolution of the gas-phase abundance of methyl cyanide and other species involved in its formation as a function of time. 
As in the case of the molecular cloud, the production of methyl cyanide requires first the formation of CH$_3$CNH$^+$, which can then be converted into CH$_3$CN. 
However, some reactions that were not important in the cold cloud, become relevant in the shock case thanks to the injection in the gas-phase of species such as methanol and ammonia, which are abundant grain mantle components (Tab. \ref{tab:model-parameters}). 
In particular, CH$_3$CN is mainly formed by the reaction of CH$_3$CNH$^+$ with ammonia (90\%) while the electron recombination of CH$_3$CNH$^+$ accounts for the $\sim$10\% of the formation of CH$_3$CN. 
In turn, 90\% of the CH$_3$CNH$^+$ is produced by the radiative association of CH$_{3}^{+}$ with HCN and the remaining 10\% by the reaction of CH$_3$OH$_2^+$ with HNC. 
In summary, the major reactions entering in the methyl cyanide formation in the shock are [7a], [9b], [10a] and [10b]. 

Finally, as for the cold molecular cloud (Sec. \ref{subsec:astro-model-cloud}) we searched for the impact of the new network on the other species of the model: they are minor and the difference in the predicted abundances before and after the network modification are less than a factor ten.

\section{Discussion} \label{sec:discussion}

Methyl cyanide is commonly observed almost ubiquitously in the ISM.
Here, we focus in particular on the solar-type star forming regions.
Table \ref{tab:detections} summarises the CH$_3$CN observations towards cold cores, low-mass protostars and hot corinos, protoplanetary disks and the only protostellar shock where it was detected, L1157-B1. 
More specifically, Table \ref{tab:detections} lists the derived CH$_3$CN column densities and abundances (with respect to H$_2$, when it is available).
It also reports the measured abundance ratio between methanol and methyl cyanide, as methanol is the most abundant iCOM.
Assuming that both species are present in the same region, their abundance ratio is a more reliable value with respect to the absolute abundance of methyl cyanide.

In the following, we first compare our results with the observations towards two sources representative of cold and warm environments (Sec. \ref{subsec:astro-compa}): TMC-1 and L1157-B1, respectively.
We then discuss the possible origin of the observed correlation between the methyl cyanide and methanol abundances, due to the chemical gas-phase link of these two species (Sec. \ref{subsec:correlation}).

\subsection{Model predictions versus observations}\label{subsec:astro-compa}

\subsubsection{Cold environment representative: TMC-1}\label{subsubsec:tmc-1}

TMC-1 (cyanopolyyne peak) is a cold starless core in the Taurus molecular cloud \citep[distance of 141 $\pm$ 9 pc;][]{zucker2019GAIA}, which shows an exceptional richness in C-bearing species.
Given its peculiarity, it has been extensively studied in several spectral surveys where carbon chains and rings were detected \citep[][and references therein]{fuente2019,cernicharo2022QUIJOTE,mcguire2020GOTHAM}.
The kinetic temperature of the gas has been estimated to be 11 $\pm$ 1 K \citep{feher2016}.
CH$_3$CN has been observed with an abundance of (with respect to H$_2$) $\sim (0.8-8)\times10^{-10}$ (see Tab. \ref{tab:detections}). 

Figure \ref{fig:model-ch3cn-abundance-cloud} shows the predicted CH$_3$CN abundance as a function of time (see Sec. \ref{subsec:astro-model-cloud}) compared to that observed towards TMC-1.
The predictions obtained with the old network slightly overproduce methyl cyanide with respect to the measured value.
On the contrary, the predictions obtained using the new network (Sec. \ref{sec:revised-network} and Tab. \ref{tab:revised-network}) agree with the lowest estimate of the measured CH$_3$CN.

Methyl cyanide is detected also towards other four cold cores: L1544 \citep[e.g.][]{Jimenez-Serra2016-L1544,Vastel2019}, L1498 \citep{Jimenez-Serra2021-L1498}, L1521E \citep{nagy2019chemical} an L1517B \citep{megias2023-PSCch3cn}.
Their abundance, which is computed assuming that the species is present throughout the entire core, has been estimated to be between $2\times 10^{-12}$ and $9\times 10^{-11}$ (see Tab. \ref{tab:detections}).
However, these abundances have to be taken as lower limits of the real ones, because, as discussed in several previous works \citep[e.g.][]{vastel2014,Ceccarelli2023-PP7}, very likely iCOMs lines in cold objects are emitted in the skins of the cores and not throughout the entire core.
Therefore, the H$_2$ column density where the iCOMs are present is smaller than that assumed, leading to a lower limit of the estimated abundance.
Considering this large uncertainty, we cannot assess if our model estimation is in agreement with the observations, nevertheless it is not in contradiction.

\subsubsection{Warm environment representative: L1157-B1}\label{subsubsec:l1157b1}

L1157-B1 is a well studied region of shocked gas, produced by the impact of a protostellar jet, driven by the Class 0 protostar L1157-mm, with the surrounding medium. 
Thanks to its rich chemistry, L1157-B1 is considered one of the best example to study the chemistry triggered by the passage of a protostellar shock, whose result is the injection of (some of) the dust grain mantles components into the gas-phase. 
Several molecules have been imaged towards L1157-B1, and ,in particular, iCOMs such as CH$_3$OH, CH$_3$CHO and NH$_2$CHO \citep{codella2017seeds, codella2020seeds}. 

To our best knowledge, L1157-B1 is the only shocked regions where methyl cyanide has been detected so far.
The images obtained with the Plateau de Bure interferometer (now called NOEMA: \url{https://iram-institute.org/observatories/noema/}) show that CH$_3$CN maps the B1 bow-shock/cavity structure, which extends over a region of about 20$"$ \citep{codella2009methyl}.
The temperature and density in this region has been the subject of several studies \citep[e.g.][]{lefloch2012ApJ,codella2020seeds,feng2022}.
Specifically, the non-LTE analysis of the methanol lines indicate a gas temperature equal to $\sim$90 K and a gas density equal to $\sim 5\times 10^{5}$ cm$^{-3}$ \citep{codella2020seeds}.
In addition, the observed [CH$_3$OH]/[CH$_3$CN] abundance ratio across the region is equal to 770--5000 \citep{codella2009methyl}. 
Using this ratio and the methanol abundance derived by \citet{codella2020seeds}, the derived CH$_3$CN abundance (with respect to H$_2$) is equal to 1.5 -- 4 $\times 10^{-6}$ \footnote{First, we obtained the methanol abundance from the methanol column density (3--8 $\times 10^{-15}$ cm$^{-2}$) derived by the non-LTE analysis of several CH$_3$OH lines towards L1157-B1 observed by \citet{codella2020seeds}.
We then divided it by the H$_2$ column density ($2 \times 10^{21}$ cm$^{-2}$) derived by \citet{lefloch2012ApJ} from the non-LTE analysis of several CO lines ($N_{CO}$ = $2\times 10^{17}$ cm$^{-2}$) and assuming the standard [CO]/[H$_2$] = $1\times 10^{-4}$ value. 
Finally, we obtained a methanol abundance (with respect to H$_2$) equal to (1.5 -- 4) $\times 10^{-6}$ and a methyl cyanide abundance of 0.3 -- 5 $\times 10^{-9}$.}.

Figure \ref{fig:model-ch3cn-abundance-shock} shows the CH$_3$CN abundance predicted by the model, described in Sec. \ref{subsec:astro-model-shock}, against the observations towards L1557-B1.
The agreement between the predictions and the observations are extremely good for shock ages larger than about 1000 yr, in line with the age of the L1157-B1 shock passage estimated by \citet{podio2017silicon}. 


\begin{figure}
	\includegraphics[width=8.5cm]{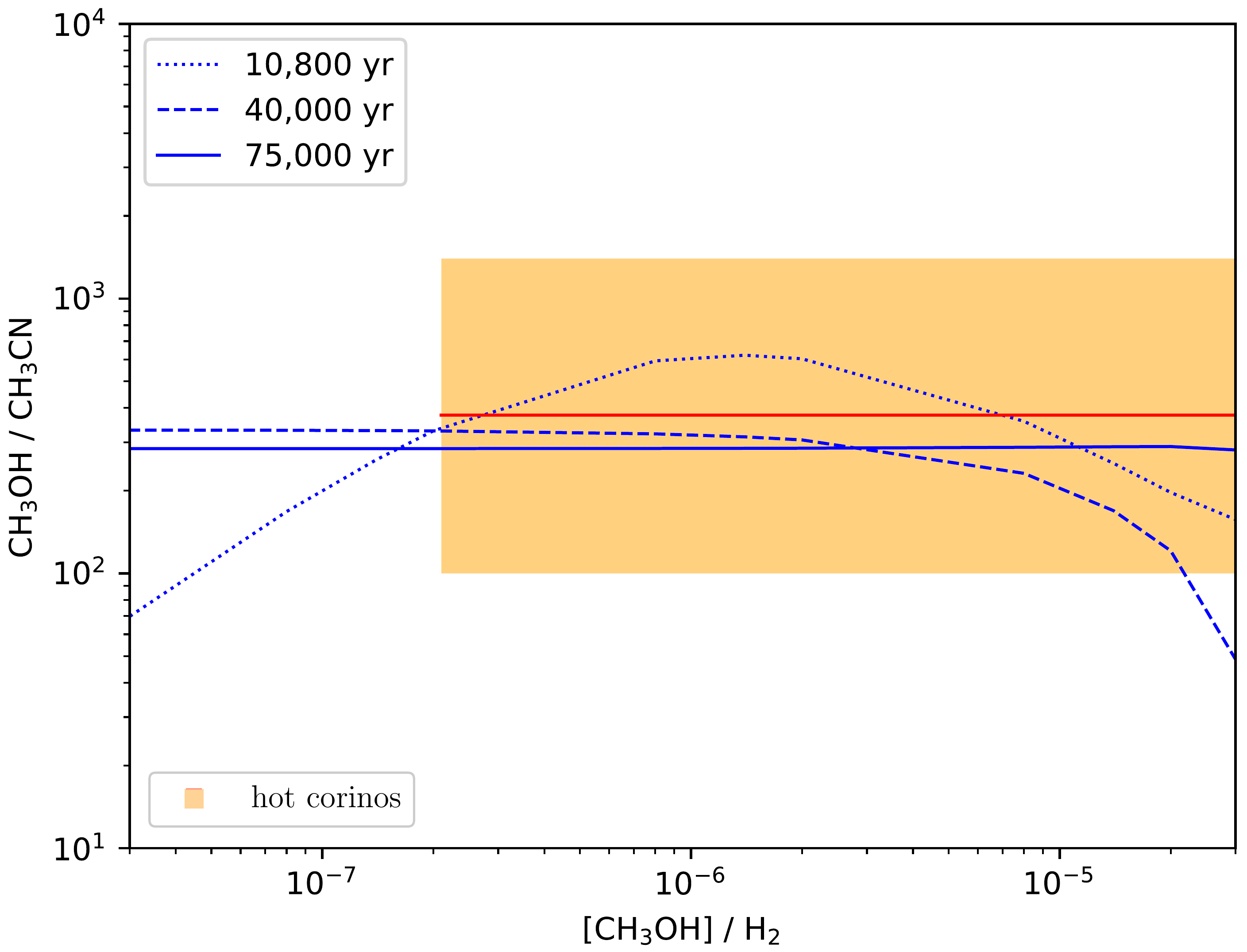}
    \caption{Comparison of observed and predicted [CH$_{3}$OH]/[CH$_{3}$CN] abundance ratio as a function of the methanol abundance (with respect to H$_2$). 
    The yellow box shows the range of the observed [CH$_{3}$OH]/[CH$_{3}$CN] ratio, derived using the values reported in Tab. \ref{tab:detections}.
    The red solid line represents the average value of the observed ratios. 
    The model predicted [CH$_{3}$OH]/[CH$_{3}$CN] ratio (blue curves) is shown for three different times: 10,800 (dotted line) , 40,000 (dashed line) and 75,000 (solid line) years, respectively. }
    \label{fig:model-ch3cn-ch3oh}
\end{figure}

\renewcommand{\arraystretch}{1.5}
\begin{table*}
    \begin{center}
	\caption{CH$_3$CN abundances measured towards cold clouds, hot corinos, protoplanetary disks and protostellar shocks} 
	\label{tab:detections}
	\begin{tabular}{lcccccc} 
		\hline
Source & N$\rm{_{CH_3CN}}$ (cm$^{-2}$) & Abundance / H$_{2}$ & N$\rm{_{H_2}}$ (cm$^{-2}$) & CH$_{3}$OH / CH$_{3}$CN & N$\rm{_{CH_3OH}}$ (cm$^{-2}$) & Reference\\
	\hline

\hline
 \multicolumn{5}{c}{\it Prestellar cores}\\
 \hline
 	\hline

TMC-1 (cyanopolyynes peak) & (4 -- 5) $\times$ 10$^{12}$ & (0.8 -- 8.0) $\times$ 10$^{-10}$ & (0.6 -- 5) $\times$ 10$^{22}$ & 7 -- 10 & (3 -- 5) $\times$ 10$^{13}$& [1,2]\\

L1544 (dust peak) & (5 -- 9) $\times$ 10$^{11}$  & (5 -- 9) $\times$ 10$^{-11}$ & 1.0 $\times$ 10$^{22}$ & 30 -- 80  & (2 -- 4) $\times$ 10$^{13}$ & [3,4]\\

L1517B (dust peak)  & (1.5 -- 2.7)  $\times$ 10$^{11}$ &  (2 -- 10) $\times$ 10$^{-12}$ & (3 -- 4)  $\times$ 10$^{22}$ & 30 -- 80 & (8 -- 12) $\times$ 10$^{12}$& [5]\\


L1498 (methanol peak) & (0.9 -- 1.1) $\times$ 10$^{11}$ & (1.3 -- 1.5) $\times$ 10$^{-11}$ & 7  $\times$ 10$^{21}$ & 80 -- 100 & 8.7 $\times$ 10$^{12}$& [6]\\

L1521E  & (3 -- 6)  $\times$ 10$^{11}$ & (2--4)  $\times$ 10$^{-11}$  & $1.6\times 10^{22}$ &  30 -- 100 & (1.7 -- 3.2) $\times$ 10$^{13}$& [7,8]\\

\hline
\multicolumn{5}{c}{\it Hot corinos}\\
 \hline
 	\hline

IRAS 16293-2422 A & (7 -- 9) $\times$ 10$^{16}$ & (2 -- 3) $\times$ 10$^{-8}$ & 4 $\times$ 10$^{24}$&  100 -- 250 & (0.9 -- 1.7) $\times$ 10$^{19}$  & [9,10,11]\\

IRAS 16293-2422 B & (3 -- 5) $\times$ 10$^{16}$ & (2 -- 4) $\times$ 10$^{-9}$ & $>$ 1.2 $\times$ 10$^{25}$ & 160 -- 400 & (0.8 -- 1.2) $\times$ 10$^{19}$ & [9,10,12]\\

SVS13-A &  (5 -- 50) $\times$ 10$^{15}$ & $<$ 5 $\times$ 10$^{-7}$ & $>$ 1 $\times$ 10$^{22}$ & 100 -- 1400 & (5 -- 7) $\times$ 10$^{18}$ &  [13,14]\\

NGC 1333 IRAS 2A  & (1.6 -- 3.2) $\times$ 10$^{16}$ & (3 -- 6) $\times$ 10$^{-9}$  & 5 $\times$ 10$^{24}$  & 100 -- 500 & (3 -- 8) $\times$ 10$^{18}$ & [15]\\

NGC 1333 IRAS 4A &  (5 -- 10) $\times$ 10$^{16}$ & (1.3 -- 2.7) $\times$ 10$^{-9}$ & 4 $\times$ 10$^{25}$ & 80 -- 440 & (0.8 -- 2.2) $\times$ 10$^{19}$ & [15]\\

B1-c &  (5 -- 10) $\times$ 10$^{15}$ & - & - & 130 -- 480 & (1.3 -- 2.5) $\times$10$^{18}$ & [16]\\

L1551-IRS5 &  $>$ 1 $\times$ 10$^{17}$  & - & $>$ 5 $\times$ 10$^{22}$ & - & $>$ 1 $\times$10$^{19}$ & [17,18]\\

HOPS-108 &  (1.8 -- 2.2) $\times$ 10$^{16}$ &  $<$ 7 $\times$ 10$^{-7}$ & $>$ 3 $\times$ 10$^{22}$ & 230 -- 780 & (5 -- 14) $\times$ 10$^{18}$ &[19]\\

Cep E-A & (0.9 -- 1.9) $\times$ 10$^{15}$ & (2 -- 5) $\times$ 10$^{-9}$ & 4 $\times$ 10$^{23}$ & 180 -- 710 & (3 -- 7) $\times$ 10$^{17}$ & [20]\\


\hline
    \multicolumn{5}{c}{\it Protoplanetary disks}\\
     \hline
 	    \hline

TW Hydrae & (1.3 -- 1.6) $\times$ 10$^{12}$ & - & - & 1.8 -- 5.0 & (3 -- 6)  $\times$ 10$^{12}$ & [21,22]\\
MWC 480 & (1.4 -- 2.2) $\times$ 10$^{12}$  & - & - & -& -& [23]\\
V4046 Sgr & (4 -- 8)  $\times$ 10$^{12}$ & - & - & -& -& [23]\\
GM Aur &  (1.9 -- 2.3)  $\times$ 10$^{12}$ & - & - & -& -& [24]\\
AS 209 & (1.5 -- 1.9) $\times$ 10$^{12}$ & - & - & -& -& [24]\\
HD 163296 & (2 -- 3) $\times$ 10$^{12}$ & - & - & -& -& [24]\\
MWC 480 & (3 -- 4) $\times$ 10$^{12}$ & - & - & -& -& [24]\\

\hline
 \multicolumn{5}{c}{\it Protostellar shocks}\\
 \hline
 	\hline
 	
L1157-B1 & (0.6 -- 10) $\times$ 10$^{12}$ & (0.3 -- 5) $\times$ 10$^{-9}$  & 2 $\times$ 10$^{21}$ & 770-5000 & (3 -- 8) $\times$ 10$^{15}$ & [see text]\\

%
%
\hline
	\end{tabular}
	\end{center}
[1] \citet{cabezas2021space};
[2] \citet{soma2015};
[3] \citet{vastel2014}; 
[4] \citet{Vastel2019};
[5] \citet{megias2023-PSCch3cn};
[6] \citet{Jimenez-Serra2021-L1498};
[7] \citet{nagy2019chemical};
[8] \citet{scibelli2021detection}; 
[9] \citet{bottinelli2004near};
[10] \citet{calcutt2018almanitriles};
[11] \citet{Manigand2020}; 
[12] \citet{Jorgensen2016}; 
[13] \citet{bianchi2022ch3cn}; 
[14] \citet{hsieh2023prodige}; 
[15] \citet{taquet2015constraining};
[16] \citet{nazari2021complex};
[17] \citet{mercimek2022chemical}; 
[18] \citet{bianchi2020faust};
[19] \citet{chahine2022organic};
[20] \citet{Ospina2018};
[21] \citet{loomis2018TWHydrae};
[22] \citet{walsh2016};
[23] \citet{bergner2018survey};
[24] \citet{ilee2021molecules};
\\
\end{table*}


\subsection{The correlation between methyl cyanide and methanol}\label{subsec:correlation}

As mentioned in Sec. \ref{sec:introduction}, a correlation between the measured abundances of CH$_{3}$CN and CH$_{3}$OH has been observed in a large sample of hot corinos \citep{belloche2020, yang2021perseus}.
Note that, in the case of \citet{yang2021perseus}, the correlation is observed between the CH$_{3}$OH and CH$_{3}$CN column densities normalised with respect to the intensity of the continuum, as a proxy of the H$_2$ column density. 
However, these authors do not explicitly derive the H$_2$ column density and, consequently, the CH$_{3}$OH and CH$_{3}$CN abundances.

The [CH$_{3}$OH]/[CH$_{3}$CN] abundance ratio is rather constant and about 380 (see Tab. \ref{tab:detections}), over a relatively large range of methyl cyanide and methanol abundances, about two orders of magnitude.
The constant ratio led \citet{yang2021perseus} to suggest the possibility that both species are formed on the grain surfaces.
On the other hand, \citet{belloche2020} ruled out that the observed methyl cyanide -- methanol correlation implies a chemical link between the two species, based on the argument that CH$_{3}$CN is mostly formed in the gas-phase and CH$_{3}$OH on the grain surfaces.
In order to understand whether the CH$_{3}$CN gas-phase and CH$_{3}$OH grain-surface formation really cannot reproduce the observed correlation, we ran the model used to reproduce the observations of L1157-B1 (Sec. \ref{subsec:astro-model-shock}) and approximately applicable also to the case of hot corinos.
Specifically, we obtained the time-dependent CH$_{3}$CN and CH$_{3}$OH predicted abundances as a function of the CH$_{3}$OH abundance.

Fig. \ref{fig:model-ch3cn-ch3oh} shows the results of the simulations along with the observations.
Please note that, in the model the methanol abundance is the one injected, which differs from that observed because methanol is consumed forming other iCOMs.
Therefore, in the plot, we consider the measured CH$_{3}$OH abundance as a lower limit to the injected one.
The figure shows the predicted [CH$_{3}$OH]/[CH$_{3}$CN] abundance ratio as a function of the methanol abundance, for different evolutionary times.
For times larger than about $4\times 10^4$ yr, the predicted [CH$_{3}$OH]/[CH$_{3}$CN] abundance ratio does not depend on the injected CH$_{3}$OH abundance, in the range of the estimated methanol abundance of the hot corinos, and it is very similar to the observed value (300--340 versus 380).
At $\sim10^4$ yr, the ratio varies within a factor 3.

The (unexpected) chemical link between methyl cyanide and methanol is the cation CH$_3^+$, which is, on the one hand, the bottleneck precursor of methyl cyanide and, on the other hand, the product of the methanol reaction with H$_3^+$. 
Specifically, methyl cyanide is mainly formed by the reaction CH$_3^+$ + HCN $\rightarrow$ CH$_3$CNH$^+$ followed by the electron recombination or proton transfer to ammonia (our reactions [7a] and [10a] of Tab. \ref{tab:revised-network}), as also discussed for the case of L1157-B1 (Sec. \ref{subsec:astro-model-shock}).
On the other hand, once methanol is injected into the gas-phase by the grain mantle sublimation, the reaction CH$_{3}$OH + H$_3^+$ $\rightarrow$ CH$_3^+$ + H$_2$ + H$_2$O is the dominant production route of CH$_3^+$.
This reaction was studied experimentally by \citet{lee1992thermal}, who measured that 53\% of the reaction goes into the channel CH$_3^+$ + H$_2$ + H$_2$O, while the protonation, that leads to the formation of CH$_3$OH$_2^+$ + H$_2$, represents only the 12\% of the products.

In summary, our simulations, obtained with the cleaned and most up-to-date gas-phase reaction network GRETOBAPE \citep{tinacci2023-gretobape},  demonstrate that the observed methyl cyanide -- methanol correlation is due to the gas-phase formation of methyl cyanide that involves the cation CH$_3^+$, which, in turn, is formed by the methanol once it is in the gas-phase.
Thus, methanol is the grandmother of methyl cyanide.

As a final and general comment, the methyl cyanide -- methanol correlation is an illustrative example of the danger to interpret the observed correlation of two species, both to assign a presumed chemical link or to exclude it.
Only a careful and dedicated modeling using our best knowledge of the chemical networks can help to assess or rule out a possible real chemical link.
We remind here the cases of methyl formate and dimethyl ether \citep{balucani2015formation}, glycolaldehyde and ethanol \citep{skouteris2018genealogical} and glycolaldehyde and acetaldehyde \citep{vazart2020gas}, all cases where the observed correlations can have a chemical-link explanation when the correct gas-phase reactions are considered.

\section{Conclusions} \label{sec:conclusions}

In this work, we present a critical review of the 15 gas-phase reactions leading to the formation of methyl cyanide invoked in the literature and reported in the two major astrochemical databases, KIDA and UDfA.
We also propose two new reactions involving protonated methanol and ammonia.
We carried out new quantum mechanics (QM) computations for eight of the 17 reactions, using a combination of DFT and CCSD(T) methods for the electronic structure calculations and RRKM theory for the kinetic. 

As a result of this study, we found that 13 of the 15 reactions reported in the KIDA and UDfA databases have incorrect products and/or rate constants.
Specifically: 
(i) we removed three reactions based on the literature review; 
(ii) we removed four  reactions based on our new QM computations; 
(iii) we modified the rate constants of one (important) reaction based on the literature search;
(iv) we modified the rate constants and products of two reactions following our new QM computations results; 
(v) we added two new reactions, whose products and rate constants were computed and compared with existing experiments;
(vi) two reactions require further investigation but, since they are not important for the methyl cyanide formation, they were not studied here.
We provide a new cleaned reaction network for the formation of methyl cyanide which consists of ten reactions, reported in Tab. \ref{tab:revised-network}.

We tested the impact of the revised network on the predicted abundance of methyl cyanide via modelling two cases, representatives of cold and warm environments: TMC-1 and L1157-B1, respectively.
First, the abundance of methyl cyanide predicted using the revised reaction network compares very well with that observed in TMC-1 and L1157-B1, respectively.
Second, we found that four reactions dominate the formation of CH$_{3}$CN (the numbers refer to those in Tab. \ref{tab:revised-network}):

\begin{tabular}{ll}
     (7a) & CH$_3^+$ + HCN $\rightarrow$ CH$_3$CNH$^+$ + $h\nu$\\
     (9b) & CH$_3$OH$_2^+$ + HNC $\rightarrow$ CH$_3$CNH$^+$ + H$_2$O\\
     (10b) & CH$_3$CNH$^+$ + e$^-$ $\rightarrow$ CH$_3$CN + H\\
     (10a) & CH$_3$CNH$^+$ + NH$_3$ $\rightarrow$ CH$_3$CN + NH$_4^+$\\
\end{tabular}

Reactions [7a] and [10b] were already present in the KIDA or UDfA databases but with incorrect rate constants, that we modified.
Specifically, for reaction [7a] we adopted a lower rate constant derived from the literature, while for reaction [10b] we adopted new calculated BFs and corrected the temperature dependence of the rate constant.  
These two reactions are confirmed to be the dominant ones in the formation of methyl cyanide in cold environments. 
Reactions [9b] and [10a] instead have been newly added to the network following the theoretical computation presented in this work, and they are found to be particularly important in warm environments.

Finally, we demonstrate that the correlation between the methyl cyanide and methanol, observed in the hot corinos where the two species are detected, can be explained by the fact that methanol is the grandmother of methyl cyanide.
The link between the two species is the cation CH$_3^+$, which is, on the one hand, the bottleneck of the reaction that dominates the CH$_3$CN gas-phase formation and, on the other hand, the product of the gas-phase reaction of H$_3^+$ with CH$_3$OH, which is a grain-surface product.

To summarise, the present work major conclusions are: 
(1) methyl cyanide can be efficiently form in the gas-phase in both cold and warm regions of the ISM, and 
(2) the correlation between methyl cyanide and methanol observed in hot corinos by other authors is likely the result of the gas-phase reaction of methanol, a grain-surface product, with H$_3^+$, which produces CH$_3^+$ which is the bottleneck for the gas-phase formation of methyl cyanide.

As final comment, we notice that, so far, gas-phase reactions can explain the observed correlations of four couples of iCOMs: 
methyl formate and dimethyl ether \citep{balucani2015formation}, glycolaldehyde and ethanol \citep{skouteris2018genealogical}, glycolaldehyde and acetaldehyde \citep{vazart2020gas} and methyl cyanide and methanol (this work).
Therefore, astronomical observations are a powerful tool to understand the chemistry of iCOMs and their correlations probably tell us of real chemical links, which is up to us to discover and which are not necessarily that the couple is formed on the grain-surfaces, as often assumed.

\section*{Acknowledgements}

This project has received funding within the European Union’s Horizon 2020 research and innovation programme from the European Research Council (ERC) for the project “The Dawn of Organic Chemistry” (DOC), grant agreement No 741002, and from the Marie Sklodowska-Curie for the project ”Astro-Chemical Origins” (ACO), grant agreement No 811312. 
The authors thank the Herla Project (http://hscw.herla.unipg.it) – “Università degli Studi di Perugia” and the “Dipartimento di Ingegneria Civile e Ambientale” of the University of Perugia within the project “Dipartimenti di Eccellenza 2018-2022” for allocated computing time. 
Some computations presented in this paper were performed using the GRICAD infrastructure (https://gricad.univ-grenoble-alpes.fr), which is partly supported by the Equip@Meso project (reference ANR-10-EQPX-29-01) of the programme Investissements d'Avenir supervised by the Agence Nationale pour la Recherche. 
E Bianchi aknowledges the Deutsche Forschungsgemeinschaft (DFG, German Research Foundation) under Germany´s Excellence Strategy – EXC 2094 – 390783311.

\section*{Data availability}
The data underlying this article are available in the article and in its online supplementary material.


\bibliographystyle{mnras}
\bibliography{CH3CN-gas-routes}


\appendix

\section{Extra Results} \label{sec:appendix-results}

\subsection{Reaction [1]: \texorpdfstring{C${_2}$H${_2^+}$ + NH$_{2}$ $\rightarrow$ CH$_{3}$CN$^{+}$ + H}{reaction1} \label{APP-subsec:results-reaction1}}

\subsubsection{PES of reaction [1]}

\begin{figure*}
	\includegraphics[width=18cm]{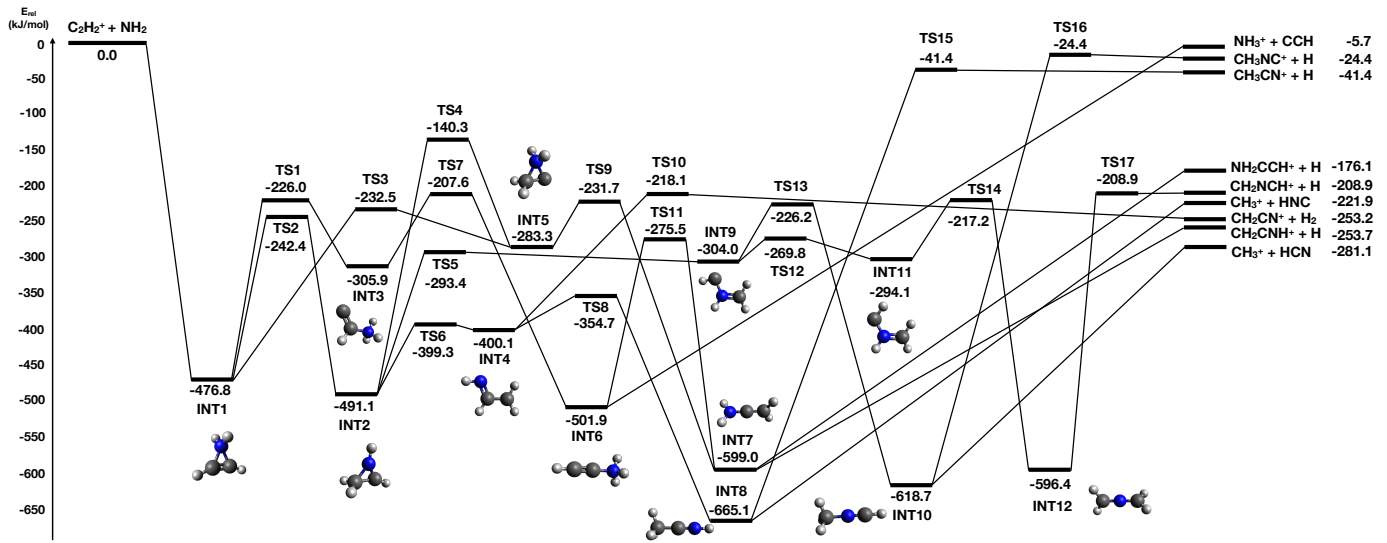}
    \caption{PES of the reaction [1] of Table \ref{tab:overview-reactionlist}, C$_{2}$H$_{2}^{+}$ + NH$_{2}$.
    The plot shows the results obtained at the CCSD(T)/aug-cc-pVTZ//B3LYP/aug-cc-pVTZ level of theory.}
    \label{fig:pes-reac1}
\end{figure*}

Figure \ref{fig:pes-reac1} shows the PES of the reaction between C${_2}$H${_2^+}$ and NH$_{2}$. Following the addition of the two reactants, the first formed intermediate is INT1, which is localised at -476.8 kJ mol$^{-1}$ with respect to the reactants. INT1 can then evolve in three different ways: it can form INT2 through TS2 (barrier of 234.4 kJ mol$^{-1}$),  INT3 through TS1 (250.8 kJ mol$^{-1}$), or it can form INT5 through TS3 (244.3 kJ mol$^{-1}$). When INT2 is formed, it can form INT4 overcoming a barrier of 91.8 kJ mol$^{-1}$ (TS6), INT9 overcoming a barrier of 197.7 kJ mol$^{-1}$ (TS5) or INT5 overcoming a barrier of 350.8 kJ mol$^{-1}$ (TS4). When INT3 is formed instead, it can isomerize to INT6 through TS7 (representing a barrier of 98.3 kJ mol$^{-1}$), and in turn INT5 can isomerize to INT7 through TS9 (whose barrier is 51.6 kJ mol$^{-1}$). Additionally, INT4  can isomerize to INT8, the most stable intermediate in the PES (located 665.1 kJ mol$^{-1}$ below the reactant energy asymptote), through TS8 (barrier of 45.4 kJ mol$^{-1}$). Otherwise INT4 can eliminate a H$_{2}$ molecule and form the CH$_{2}$CN$^+$ + H$_{2}$ products, localised at -253.2 kJ mol$^{-1}$ with respect to the reactants. INT6 can instead isomerize to INT7 through TS11 (barrier of 226.4 kJ mol$^{-1}$) or it can go through a proton transfer process and form NH$_{3}^{+}$ + CCH, localised at -5.7 kJ mol$^{-1}$ with respect to the reactants. INT9 can isomerize to INT10 via TS13 (77.8 kJ mol$^{-1}$) or into INT11 via TS12 (34.2 kJ mol$^{-1}$), while INT11 in turn can evolve into INT12 through TS14 (barrier of 76.9 kJ mol$^{-1}$). INT7 can lose an H-atom and form NH$_{2}$CCH$^+$ + H or CH$_{2}$CNH$^+$ + H, respectively. The fission of the N-H bond in INT8 leads to the formation of CH$_{3}$CN$^{+}$ + H trough TS15 (628.0 kJ mol$^{-1}$), otherwise the fission of the C-C bond leads to the formation of CH$_{3}^{+}$ + HNC in a barrier-less process. The same fate of INT8 can be observed starting from the INT10 intermediate, which can form CH$_{3}$NC$^{+}$ + H through TS16, or CH$_{3}^{+}$ + HNC without any barrier. Finally, the fission of the C-H bond in INT12 (TS17) can form CH$_{2}$NCH$^+$ + H localised at -208.9 kJ mol$^{-1}$. We remark that TS15, TS16 and TS17 were localised as submerged transition states and therefore are reported with the same energy of the respective products.

\subsubsection{Kinetics of reaction [1]}

For this system we didn't run the RRKM calculations but we noticed that the channel of formation of CH$_{3}$CN$^{+}$ has probably a negligible BF with respect to the other channels present in the PES. The channel of formation of CH$_{3}$CN$^{+}$ is indeed much less exothermic (-41.4 kJ mol$^{-1}$) than the other competitive channels (up to -281.1 kJ mol$^{-1}$). Moreover, the formation of CH$_{3}$CN$^{+}$ presents high energy barriers along the reaction pathways: INT8 can lose an H atom and form CH$_{3}$CN$^{+}$ only overcoming a barrier of 628.0 kJ mol$^{-1}$, while the competitive process, the dissociation of INT8 into CH$_{3}^{+}$ and HNC, is barrier-less and leads to the formation of more stable products (-221.9 kJ mol$^{-1}$). 
For the proton transfer process we could not find a direct channel connecting the reactants to the products via a single adduct. The NH$_{3}^{+}$ + CCH products can be formed in a multi-step process that requires the formation of INT1, INT3 and then INT6. It seems therefore that this reaction can not be an important source of CH$_{3}$CN$^{+}$ as reported in the databases and we decided to remove it from our network. RRKM calculation to derive the specific BFs are postponed to a future work.

\subsection{Reaction [2]: \texorpdfstring{C${_2}$H$_{4}^{+}$ + N  $\rightarrow$ CH$_{3}$CN$^{+}$ + H}{reaction2} \label{APP-subsec:results-reaction2}}

\subsubsection{PES of reaction [2]}

Figure \ref{fig:pes-reac2} shows the full reaction path following the reaction of N atom with the C${_2}$H$_{4}^{+}$ ion in their fundamental states ($^4$S) and ($^2$A), respectively. The first intermediate formed is INT1 in the triplet state which can isomerize to INT2, INT3 or INT6 through a barrier of 194.0 kJ mol$^{-1}$ (TS1), 48.9 kJ mol$^{-1}$ (TS2) and 223.7 kJ mol$^{-1}$ (TS3), respectively. INT2 can in turn isomerize to INT5 via TS4 overcoming a barrier of 71.6 kJ mol$^{-1}$ or into INT4 overcoming a barrier which is localized at 0.72 kJ mol$^{-1}$ above the energy of the reactants (TS5). INT3 can instead isomerize to INT6 through a barrier of 243.7 kJ mol$^{-1}$ (TS6), or it can lose an H atom an form CH${_2}$NCH$^{+}$, the second most exothermic product (-220.3 kJ mol$^{-1}$ with respect to the reactants), by overcoming a barrier of 154.7 kJ mol$^{-1}$ (TS7). INT4 can isomerize to INT8 through a barrier of 67.8 kJ mol$^{-1}$ (TS9), while INT5 can form INT9 through TS11 (representing a barrier of 228.1 kJ mol$^{-1}$) or INT8 through TS8 (barrier of 0.7 kJ mol$^{-1}$). INT6 can isomerize to INT7 through TS10 (barrier of 70.2 kJ mol$^{-1}$) and both of them can lose an H atom in order to form CH$_{3}$NC$^{+}$ (localised at -35.8 kJ mol$^{-1}$ with respect to the reactants), and CH${_2}$NCH$^{+}$ (localised at -52.7 kJ mol$^{-1}$ with respect to the reactants), respectively. INT8 can instead form CH${_2}$CNH$^{+}$, the most exothermic product (-265.1 kJ mol$^{-1}$ with respect to the reactants), overcoming a barrier of 131.8 kJ mol$^{-1}$ (TS16). INT8 can also isomerize to INT10 through TS15 (barrier of 201.1 kJ mol$^{-1}$), as INT9 can isomerize to INT10 through TS14 (barrier of 31.1 kJ mol$^{-1}$). INT10 can in turn isomerize to INT11 overcoming a barrier of 151.8 kJ mol$^{-1}$ (TS19), otherwise it can lose an H atom and form CH${_2}$CNH$^{+}$ through TS18 (barrier of 72.4 kJ mol$^{-1}$). INT7 can also lose an H-atom and form CH$_{3}$CN$^{+}$ by overcoming a barrier of 190.0 kJ mol$^{-1}$ (TS20).

The possibility of ISC from the triplet surface to the more stable singlet surface has not been explored at this level of the computations since our aim for this reaction was to check whether the product is CH${_2}$CNH$^{+}$, as reported by \citet{scott1999c}, or CH$_{3}$CN$^{+}$, as reported by \citet{loison2014interstellar}. The presence of ISC would not turn CH$_{3}$CN$^{+}$ into the main product of the reaction, therefore we focused our calculations only on the triplet surface. Moreover only the products that retain the CCN chain are shown while all the channels that require the breaking of the C-C or the C-N bond are neglected. A complete investigation of the reaction is postponed to a possible future work.

\subsubsection{Kinetics of reaction [2]}

In this reaction the most favourable product is CH${_2}$NCH$^{+}$, since in can be formed in a two step process and its channel of formation is highly exothermic (-220.3 kJ mol$^{-1}$). All the other products CH$_{3}$NC$^{+}$, CH${_2}$CNH$^{+}$ and CH$_{3}$CN$^{+}$ require at least four steps to be formed. Given that CH$_{3}$CN$^{+}$, the product reported in the databases, is much less exothermic than the two isomers CH$_{3}$NC$^{+}$ and CH${_2}$CNH$^{+}$, and that it requires more steps than CH$_{3}$NC$^{+}$ to be formed, it should not be considered the reaction's major product. Because we were solely interested in the formation of CH$_3$CN$+$, we chose to eliminate this reaction from our network without doing kinetic calculations on the PES.

However, the formation of CH${_2}$CNH$^{+}$ and CH${_2}$NCH$^{+}$ is of relevance since the CH${_2}$CN radical has been observed in the ISM \citep{irvine1988identification} and it might be linked to CH$_{3}$CN. More investigations are needed but they are postponed to a future work.

\begin{figure*}
	\includegraphics[width=18cm]{figures/PES-reac2-c2h4p-n.pdf}
    \caption{PES of the reaction [2] of Table \ref{tab:overview-reactionlist}, C$_{2}$H$_{4}^{+}$ + N.
    The plot shows the results obtained at the CCSD(T)/aug-cc-pVTZ//B3LYP/aug-cc-pVTZ level of theory.}
    \label{fig:pes-reac2}
\end{figure*}

\subsection{Reaction [4]: \texorpdfstring{H${_2}$CCN + H${_3^+}$ $\rightarrow$ CH$_{3}$CN$^{+}$ + H$_{2}$}{reaction4} \label{APP-subsec:results-reaction4}}

\subsubsection{PES of reaction [4]}

\begin{figure}
	\includegraphics[width=8.5cm]{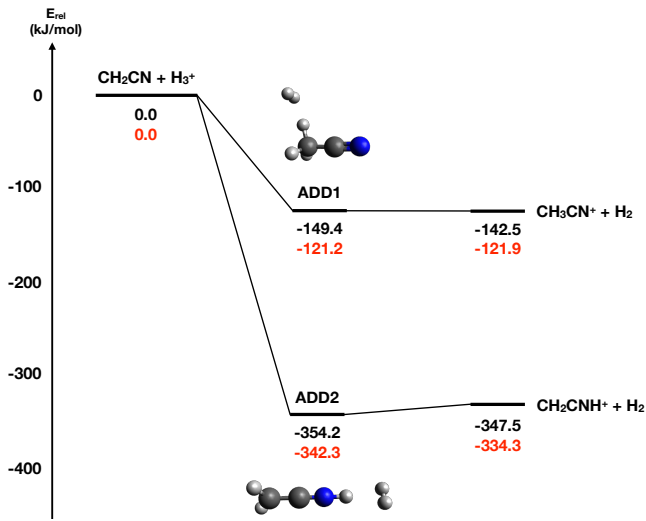}
    \caption{PES of the reaction [4] of Table \ref{tab:overview-reactionlist}, CH$_{2}$CN + H$_{3}^{+}$. In the plot the B3LYP/aug-cc-pVTZ energies are reported in black, while the CCSD(T)/aug-cc-pVTZ energies are reported in red.}
    \label{fig:pes-reac4}
\end{figure}

The cyanomethyl radical CH${_2}$CN exists in two resonant forms, the first one with a triple bond between the C and N atoms and the unpaired electron on the terminal carbon atom, and the second one with both the C-C and C-N atom pairs connected by double bonds and the unpaired electron on the N atom. Since the form with the radical on the N atom is the favoured one, we expect the protonation by H$_{3}^{+}$ to be more favourable on the N atom than that on the C atom. This is confirmed by the PES showed in \ref{fig:pes-reac4}: the reactants can form two adducts which are almost isoenergetic with the products CH$_{3}$CN$^{+}$ ($\Delta H$= -121.9 kJ mol$^{-1}$) and CH${_2}$CNH$^{+}$ ($\Delta H$= -334.2 kJ mol$^{-1}$), respectively. In particular ADD1 is localised at -6.9 kJ mol-1 below the energy of the products at B3LYP level, but it disapper at CCSD(T) level, while ADD2 is localised below the products both at CCSD(T) and B3LYP level, at -8.0 kJ mol$^{-1}$ and -6.7 kJ mol$^{-1}$ respectively. Once the adducts are formed, they can eliminate an H$_2$ molecule and from the products without any barrier.

\subsubsection{Kinetics of reaction [4]}

Considering that the CH${_2}$CN most abundant form is the one with the radical on the N atom, and that the channel of formation of CH${_2}$CNH$^{+}$ is 213 kJ mol$^{-1}$ more exothermic than the channel of formation of CH$_{3}$CN$^{+}$, we concluded that the product of the reaction is mainly CH${_2}$CNH$^{+}$ and not CH$_{3}$CN$^{+}$ as reported in the databases. We therefore decided to remove the reaction from our network but, as for reaction [2], the formation of CH${_2}$CNH$^{+}$ is still of interest and opens new questions on the connection between CH$_{3}$CN and CH${_2}$CN. 

\subsection{Reaction [8]: \texorpdfstring{CH$_{4}^{+}$ + HCN $\rightarrow$ CH$_{3}$CNH$^{+}$ + H}{reaction8} \label{APP-subsec:results-reaction8}}

\subsubsection{PES of reaction [8]}
\begin{figure*}
	\includegraphics[width=16cm]{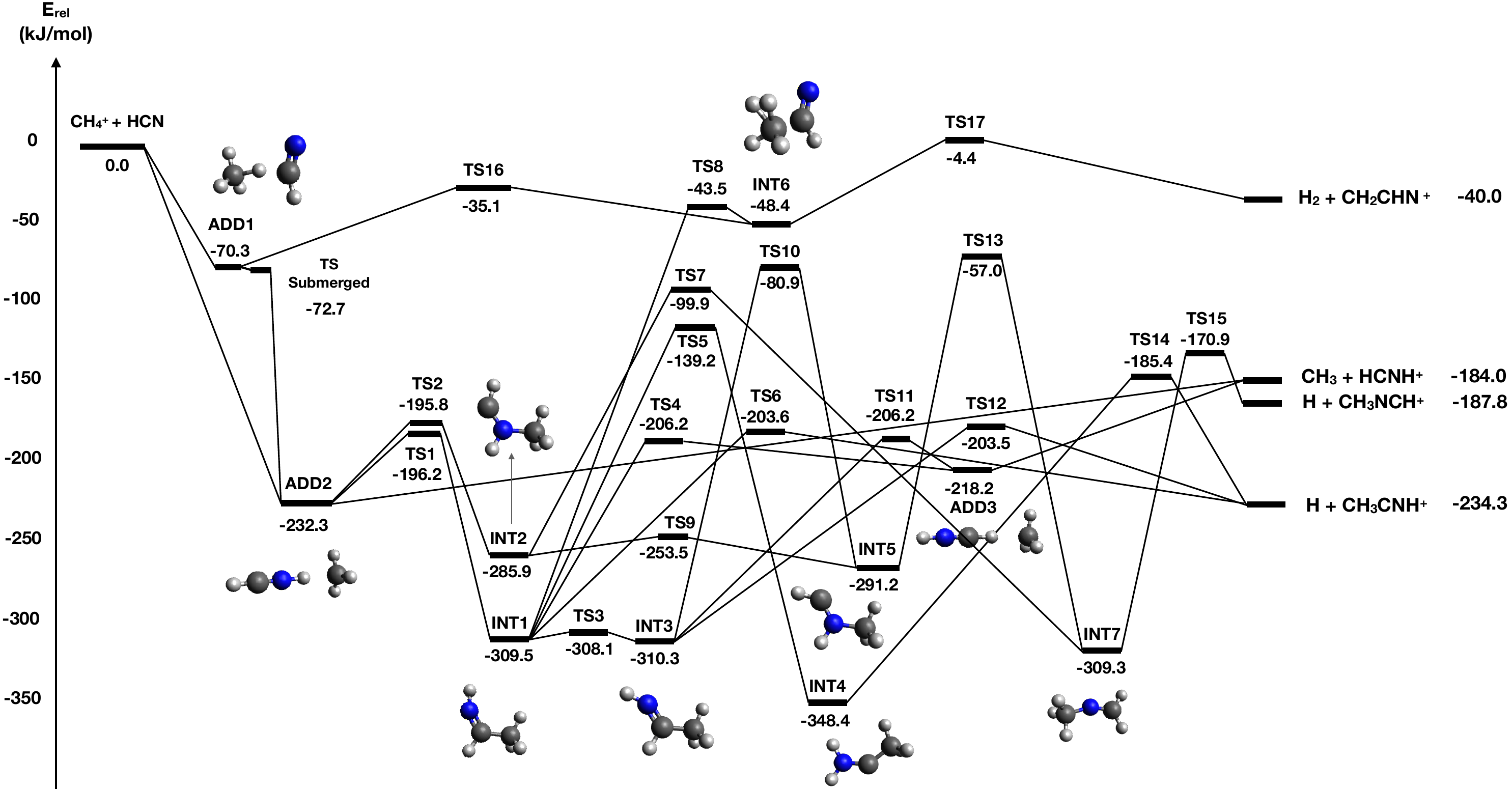}
    \caption{PES of the reaction [8] of Table \ref{tab:overview-reactionlist}, CH$_{4}^{+}$ + HCN.
    The plot shows the results obtained at the CCSD(T)/aug-cc-pVTZ//B3LYP/aug-cc-pVTZ level of theory.}
    \label{fig:pes-reac8}
\end{figure*}

Figure \ref{fig:pes-reac8} shows the full reaction path following the barrier-less addition of HCN on the ion CH$_{4}^{+}$. 
Starting from the reactants, there are two adducts that can be formed without any barrier, ADD1 and ADD2, localised at -70.3 and -232.3 kJ mol$^{-1}$ with respect to the reactants, respectively. 
However, the lifetime of ADD1 is so short that when it is formed it converts directly into ADD2 through a saddle point that is localised at B3LYP level, but not at CCSD(T) level. 
For this reason, ADD1 is not relevant for the kinetic of the reaction, but it is reported as a stationary point in the PES.

Once ADD2 is formed, a proton transfer can occur to form the CH$_3$ + HCNH$^+$ products, which are localized at -184.0 kJ mol$^{-1}$ with respect to the reactants.
Alternatively, it can form two different intermediates, INT1 or INT2 through a barrier of 36.1 kJ mol$^{-1}$ (TS1) and 36.5 kJ mol$^{-1}$ (TS2), respectively. 
INT1 can isomerize to four different minima: INT3, overcoming a barrier of 1.4 kJ mol$^{-1}$ (TS3), INT4 through a barrier of 170.3 kJ mol$^{-1}$ (TS5), INT6 through a barrier of 266.0 kJ mol$^{-1}$ (TS8), or ADD3 through a barrier of 103.3 kJ mol$^{-1}$ (TS4). 
Additionally, INT1 can form the CH$_3$CNH$^+$ + H products, localized 234.3 kJ mol$^{-1}$ below the reactant energy asymptote, overcoming a barrier of 105.9 kJ mol$^{-1}$ (TS6). 
When INT2 is formed from ADD2, it can isomerize to INT5 via TS9 (barrier of 32.4 kJ mol$^{-1}$) or INT7 via TS7 (barrier of 186.0 kJ mol$^{-1}$). 
INT3 can instead isomerize to INT5, through TS10 (barrier of 229.4 kJ mol$^{-1}$) or ADD3 through TS11 (barrier of 104.1 kJ mol$^{-1}$), otherwise it can form CH$_3$CNH$^+$ + H overcoming a barrier of 106.8 kJ mol$^{-1}$ (TS12). 
Also INT4 can form CH$_3$CNH$^+$ + H through TS14 (163 kJ mol$^{-1}$ of barrier), while INT7 can form CH$_3$NCH$^+$ + H, localized at -187.8 kJ mol$^{-1}$, overcoming a barrier of 138.4 kJ mol$^{-1}$ (TS15). 
ADD3 can undergo a barrier-less proton transfer process as ADD2 and form CH$_3$ + HCNH$^+$. 
ADD1 can also isomerize to INT6 through TS16 (-35.1 kJ mol$^{-1}$ with respect to the reactants), which in turn can form CH$_{2}$CNH$^+$ + H$_{2}$, the less exothermic products of the reaction (-40.0 kJ mol$^{-1}$), overcoming TS17 localized at -4.4 kJ mol$^{-1}$. 
Our work is in agreement with the one by \citet{li2008theoretical}, however some new transition state (TS12) and intermediates (INT5 and ADD3) have been found, as well as a connection between the reactants and ADD1.

\subsubsection{Kinetics of reaction [8]}

We computed the entrance potential using the electronic structure calculations described in Sec. \ref{subsec:methods-kinetics} and we derived a capture rate constant of $3.9 \times 10^{-9}$ cm$^3$ s$^{-1}$. 
This value has been used to calculate the rate constants of formation of each product:
$3.9 \times 10^{-9}$ cm$^3$ s$^{-1}$ for the proton transfer channel that forms CH$_3$ + HCNH$^+$;
$3.2 \times 10^{-11}$ cm$^3$ s$^{-1}$ for the channel of formation of CH$_3$CNH$^+$ + H; and $3.0 \times 10^{-15}$ cm$^3$ s$^{-1}$ for the channel of formation of CH$_3$NCH$^+$ + H. 
The BF of the channel of formation of CH$_{2}$CHN$^+$ + H$_{2}$ is negligible with respect to the other three. 

\begin{figure}
    \includegraphics[width=8cm]{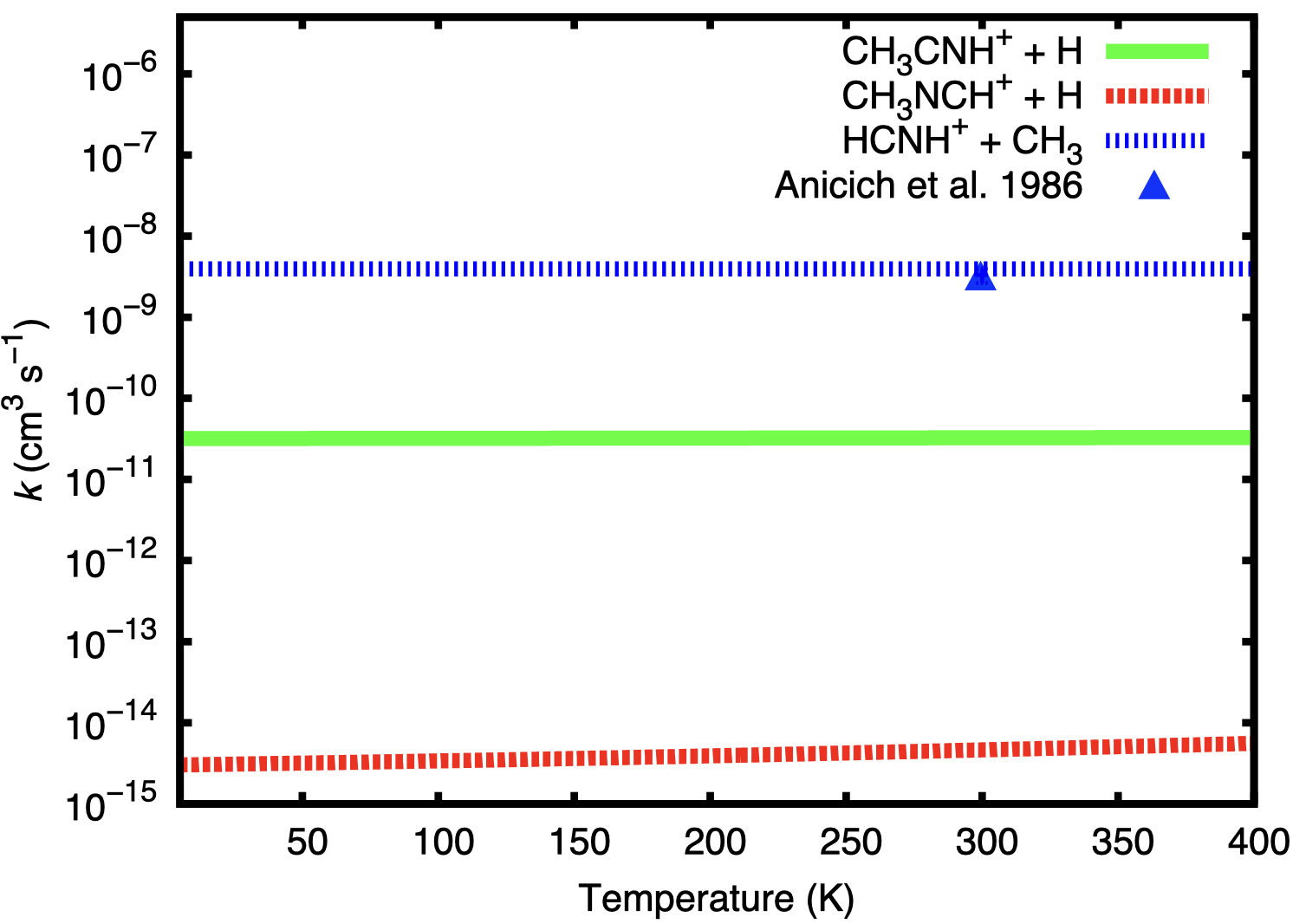}
    \caption{Rate constants of reaction [8], CH$_{4}^{+}$ + HCN, as a function of the temperature for the products CH$_3$ + HCNH$^+$ (blue), CH$_3$CNH$^+$ + H (green) and CH$_3$NCH$^+$ + H (red), respectively.
    The plot also reports the measurements at room temperature of the rate constant of the CH$_3$ + HCNH$^+$ product \citep{bass1981ion,anicich1986ion}. }
    \label{fig:rate-reac8-9b}
\end{figure}

The obtained rate constants as a function of the temperature are shown in Fig. \ref{fig:rate-reac8-9b}. 
There is no dependency on the temperature, which is expected for ion-molecule reactions, and the back dissociation is negligible since the first intermediate that is formed is highly stable (-232.3 kJ mol$^{-1}$). 

Finally, our computed rate constant at 300 K for the proton transfer channel forming CH$_3$ + HCNH$^+$ agrees very well with the two experimental measurements by \cite{bass1981ion} and \cite{anicich1986ion}.

\bsp	
\label{lastpage}
\end{document}